\documentclass[footinbib,twocolumn,showpacs,amsmath,amstex,amssymb,mathfonts,superscriptaddress,prx]{revtex4}
\usepackage{graphicx}
\usepackage{color}
\usepackage{bm}

\usepackage{graphicx}
\usepackage{color}
\usepackage{bm}
\usepackage{amsmath}
\usepackage{amssymb}
\usepackage{amsthm}
\usepackage{amsfonts}
\usepackage{multirow}
\usepackage{makecell}
\usepackage{float}
\usepackage{comment}
\usepackage{enumitem}
\usepackage{verbatim}
\usepackage{mathtools}

\begin{document}

\title{Gaplessness protected by bulk-edge correspondence}
\author{Yoshiki Fukusumi} 
\affiliation{Division of Physics and Applied Physics, Nanyang Technological University, Singapore 637371.}

\date{\today}
\begin{abstract}
After almost half a century of Laughlin's celebrated study of the wavefunctions of integer and fractional quantum Hall (FQH) effects, there have still existed difficulties to prove whether the given wavefunction can describe a gapped phase or not in general. In this work, we show the FQH states constructed from nonunitary conformal field theories (CFTs), such as Gaffiinian and Haldane-Rezayi states have difficulty gapping out under preserving bulk-edge correspondence in the cylinder geometry. Contrary to the common understandings of the condensed matter communities, the gaplessness for these systems seems not to come from the negative conformal dimensions of nonunitary CFTs in this setting at least directly. We propose the difficulty is coming from the mismatch of the monodromy charge of a quantum state and the simple charge of the corresponding operator in underlying CFTs, known as Galois shuffle. In the Haldane-Rezayi state, this corresponds to the conjugate operation of the Neveu-Schwartz and Ramond sectors for the unitary Weyl fermion and the nonunitary symplectic fermion. In the Gaffinian state, besides the Galois shuffle structure, the anomalous conformal dimension of the $Z_{2}$ simple current results in the cylinder partition functions outside of the existing local quantum field theory. This indicates the existing gapless fractional quantum Hall states have similar nonlocal structures, similar to deconfined quantum criticality. Our work opens up a new paradigm which gives a criterion to predict whether the candidate of topologically ordered states is gapped or not, and local or nonlocal, by revisiting the problem of anomaly and the duality of symplectic and Dirac fermion. 

\end{abstract}

\maketitle 

\section{Introduction}
\label{sec:introduction}

Recently, the modular properties of a conformal field theory (CFT) under the application of symmetry action have captured some attention of condensed matter physicists to consider its realization in one dimensional quantum spin chain\cite{Furuya:2015coa,Numasawa:2017crf,Kikuchi:2019ytf}. In these works, the modular property of the orbifolded theory and the consistency conditions of the boundary state have been discussed\cite{Han:2017hdv}. One may be possible to relate this correspondence to the recent proposal by Cardy which conjectures the correspondence between renormalization group (RG) flow and boundary states\cite{Cardy:2004hm,Lencses:2018paa}. However, as was stressed by older works by mathematical physicists and high energy physicists, its "physical" meaning of modular invariance and the existence of BCFT are not clear when considering two dimensional statistical mechanical systems or one dimensional quantum systems\cite{Gannon:2001ki}. A typical example in the recent condensed matter physics is the parafermionic chain in the closed and open boundary conditions\cite{Yao:2020dqx}. As was shown in \cite{Yao:2020dqx}, the parafermionic representation of the parafermionic chain does not satisfy the usual modular properties when considering the CFT description. However, there exists a mapping of boundary states in open chain geometry by the Fradkin-Kadanoff transformations\cite{Fendley:2012vv,Fukusumi:2020irh}. Hence this implies the existence of BCFTs which do not correspond to modular invariant and indicates a difficulty to use the modular property as a guiding criterion to distinguish quantum phases of matter in one dimensional qunatum systems\footnote{A possibly related arugument about the modular properties of 2d statistical mechanical models can be seen in \cite{Jacobsen:2022nxs}.}.

\begin{figure}[htbp]
\begin{center}
\includegraphics[width=0.5\textwidth]{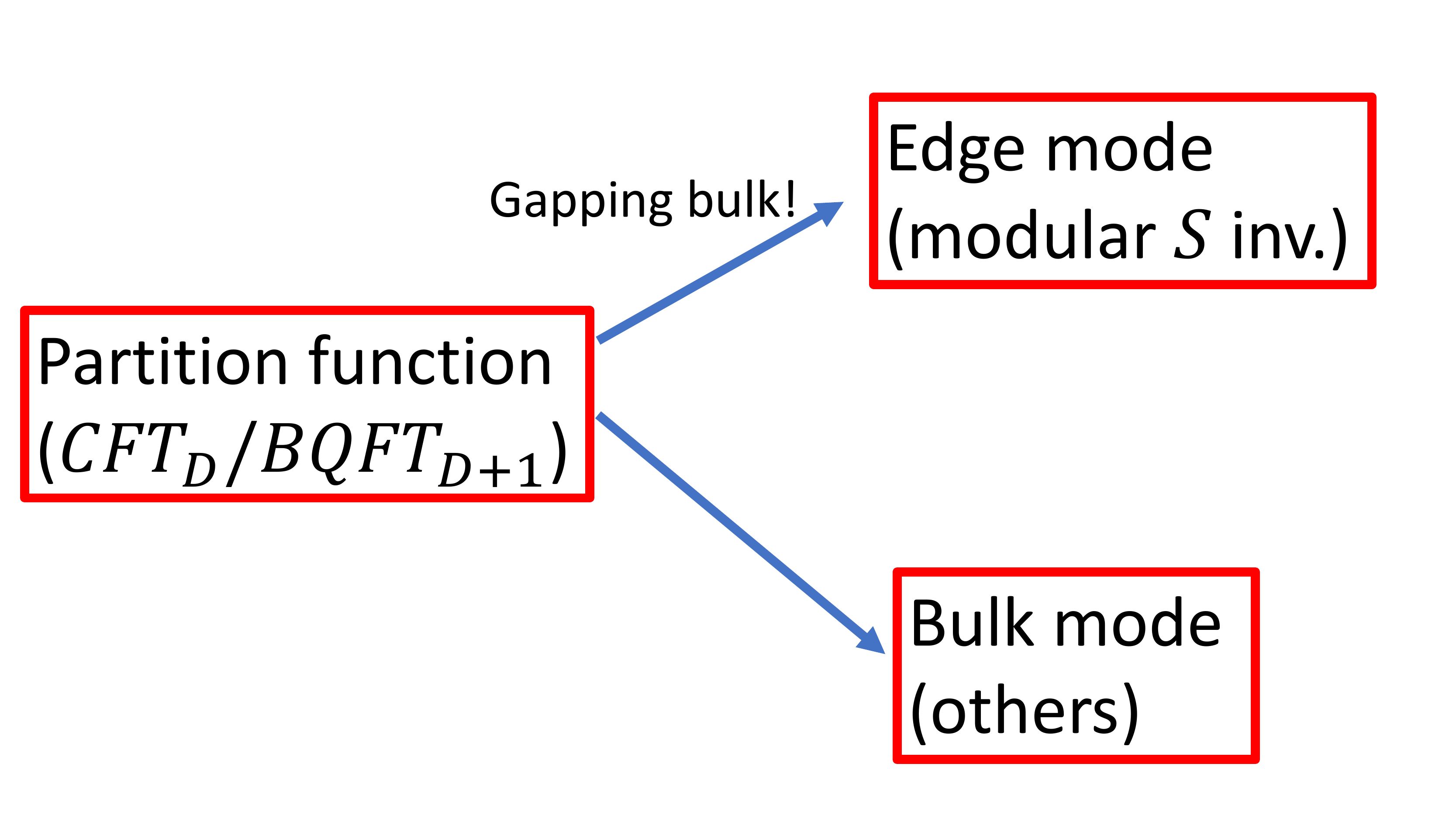}
\caption{RG flow of CFT/BQFT correspondence. The existing lierature implies a rostness of modular $S$ invariant part of the theory, which corresponds to the Lagrangian subalgebra}\label{modular}
\end{center}
\end{figure}

Instead, one of the promissive approaches to consider the modular property of the two dimensional CFTs is in the context of topologically ordered systems and the fractional quantum Hall effect (FQHE). The significance of the modular property for gapped FQHE was first recognized by Cappelli and his collaborators \cite{Cappelli:1996np,Cappelli:1998ma,Cappelli:2013iga}, and they have clarified its physical meaning in a class of condensed matter systems,  $SU(m)\times U(1)$ model and related $W_{1+\infty}$ model. There also exists analysis for the general topological ordered systems with bulk-edge correspondence \cite{Park:2016xhc,Chen_2016}. However, the analysis of the modular property of FQHE systems is only limited to such a particular class of systems, whereas the general construction of wavefunction of FQHE has already appeared\cite{Schoutens:2015uia}. Moreover, the meaning of modular noninvariance in gapless FQHE has rarely been considered, except for Haldane-Rezayi state\cite{Moore:1991ks,Ino:1998by}. More recently, the modular $S$ invariance of cylinder partition functions has been proposed as a necessary condition to obtain stable gapped edges in the general topological ordered states\cite{2013PhRvX...3b1009L}, and related proposals appeared\cite{Chatterjee:2022tyg,Chatterjee:2022kxb}. Based on the identification of bulk-edge correspondence as between $D+1$ dimensional massless quantum field theory with boundary and $D$ dimensional CFT ($CFT_{D}/BQFT_{D+1}$ correspondence) where the spacetime dimension is $D+1$, one can understand the appearance of edge modes under bulk gapping operation as a consequence of the modular $S$ invariant part of the total partition function\cite{Fukusumi_2022_f}(Fig.\ref{modular}).
Related proposal for this correspondence in the language of AdS/CFT or Lagrangian formalism, one can see \cite{Nguyen:2022khy,Ino:2000wa} and \cite{Jolicoeur:2014isa,Estienne2015CorrelationLA} for numerical results. Hence, to check the validity of the modular invariance as a general stability criterion for the topological phases of matter, it is a good starting point to study the modular property of other classes of gapless and gapped FQHE systems.

In this work, we show several modern aspects of nonunitary FQHE models by considering the quantum anomaly and Galois shuffle and their implication in FQHE\cite{Gannon:2003de}. These FQH models constructed from nonunitary CFTs have been conjectured to become gapless and to correspond to quantum critical models\cite{Estienne2015CorrelationLA,Jolicoeur:2014isa,Weerasinghe_2014}. There exist theoretical proposals which indicate the gaplessness of these models by considering the Lagrangian description of the chiral edge modes and braiding relations\cite{Read:2007cv,Read:2008rn}. However, it is difficult to show this conjecture because of the size limitation of lattice models numerically. Moreover, even analytically, there also exist some difficulties to consider chiral edge modes as protected edge modes, so we consider cylinder geometry and its modular property. Our analysis is based on the recent proposal of anomaly free fermionic CFT and older proposal of "anomalous" fermionic objects which appear in the partition functions\cite{Kedem:1993ze,Berkovich:1994es,Jacob:2005kc,Hsieh:2020uwb,Kulp:2020iet,Lou:2020gfq}. There also exists a proposal to construct the anomalous fermionic theory by using category theory\cite{Inamura:2022lun}.

The rest of the paper is organized as follows. In section \ref{u1}, we introduce the notations and the basic construction of the FQH wavefunction briefly. In section \ref{fermion}, which is the main part of this work, the difficulty of gapping for Gaffnian states under the bulk-edge correspondence is shown. In section \ref{HR}, we revisit the gaplessness of Haldane-Rezayi states, with emphasis on anomaly and Galois shuffle. In section \ref{conclusion}, we make a concluding remark. In appendix, we show detailed calculations related to anomalous gauging or fermionization and its implication for the lattice models.

\section{$U(1)$ part and modular property}
\label{u1}

Let us introduce the modular $S$ and $T$ transformations,
\begin{align} 
S:& \tau\rightarrow -\frac{1}{\tau},\\
T:& \tau \rightarrow \tau+1,\\
\end{align}
where the modular parameter $\tau$ is related to the partition function as $Z(\tau)=\text{Tr} e^{-2\pi i \tau H_{CFT}}$. In the system with bulk-edge correspondence in $2+1$ dimension, we identify the spectrum of the total system as  $H_{CFT}$. After gapping bulk, one can expect to observe edge modes of the system which are usually described by Lagrangian or local CFT with modular $S$ and $T^{2}$ invariance.

It should be noted that the $U(1)$ chiral Luttinger part of the FQHE system is often omitted from the analysis because it is not necessary to take into account when calculating several physical quantities such as topological entanglement entropy. However, it is not the case for our study. For example, when we restrict our attention to $c=-2$ CFT corresponding to the Haldane-Rezayi state, we can easily a obtain modular invariant partition function when we neglect the contribution from $U(1)$ part\cite{Creutzig:2013hma,Guruswamy:1996rk}. The same is true for other nonunitary CFTs which appears as Jack polynomial in FQHE\cite{Bernevig_2008,Bernevig_2009}. Hence the modular property without considering the contribution of $U(1)$ part is not a sufficient criterion to distinguish gapless and gapped FQHE states.
In this work, we concentrate on modular $S$ and modular $T$ property of the model, because the models in our scope do not satisfy modular $S$ and $T$ invariance as the necessary condition to obtain the protected edge modes. There exist other modular transformations, denoted as $U$ and $V$ which are relted to flux property of generalized Gibbs ensambre constructed from the edge spectrum\cite{Cappelli:1996np}.

Then, let us introduce the partition function of the Laughlin states. The chiral partition function of the Laughlin states is identified as the chiral character of $U(1)$ Wess-Zumino-Witten model,
\begin{equation}
\theta_{\frac{r}{q}}^{+}=\frac{1}{\eta(x)}\sum_{m=-\infty}^{\infty} x^{\frac{\left(2qm+r\right)^{2}}{2q}}, \ x=e^{2\pi i\tau},
\end{equation}
where $q$ is an half-integer and $r$ corresponds to quasihole excitations, $r=0,..,2q$ and $\eta(x)=x{-1/24}\prod_{n=1}^{\infty}(1-x^{n})$ is the eta function, and $m$ corresponds to particle counting or (fermionic) parity.

For the later discussion, we introduce a character as in \cite{Milovanovic:1996nj,Ino:1998by},
\begin{equation}
\theta_{\frac{r}{q}}^{-}=\frac{1}{\eta(x)}\sum_{m=-\infty}^{\infty}\left(-1\right)^{m} x^{\frac{\left(2qm+r\right)^{2}}{2q}}
\end{equation}
Also, we introduce the following parity even and odd characters,
\begin{equation}
\theta_{\frac{r}{q}}^{\text{even}}=\frac{1}{2}\left(\theta_{\frac{r}{q}}^{+}+\theta_{\frac{r}{q}}^{-}\right),\ \theta_{\frac{r}{q}}^{\text{odd}}=\frac{1}{2}\left(\theta_{\frac{r}{q}}^{+}-\theta_{\frac{r}{q}}^{-}\right).
\end{equation}

Here, we introduce the $Z_{2}$ simple charge of operators to couple a general CFT and the Laughlin states. The $Z_{2}$ simple charge of the operator indexed by $\alpha$ is,
 \begin{equation}
Q_{J}(\alpha)=h_{\alpha}+h_{J}-h_{J(\alpha)},
\end{equation}
where $J$ is the $Z_{2}$ simple current and $h$ is the conformal dimension of the respective fields. This defines the twist of an operator as $Q_{J}(\alpha)=\text{half-integer}$.

The most crucial point to consider the FQH wavefunctions in CFT is the singlevaluedness of the wavefunctions. By introducing the new electron operator after coupling the CFT and Laughlin states as $\Psi=Je^{i\sqrt{q}\phi}$, this condition changes the structure of Laughlin states as $r\rightarrow r+1/2$ for the twisted sector. In the following section, we fix $r$ as an integer and represent the character corresponding to half flux quantum sector by redefinition of the index in this way. As we will see, this changes the modular property of partition functions entirely.

In the quantum spin chain, this half-flux quantum attachment corresponds to assigning the antiperiodic boundary condition and it is closely related to fermionization \cite{Alcaraz:1987zr,Kitazawa_1997,Nomura:1997pp,Thorngren:2018bhj,Yao:2019bub}. Similar modification appears in the analysis of geometric clusters in the two dimensional statistical models, which are outside of the Kac table and idetifications\cite{Picco:2016ilr,Gori:2018gqx,Picco:2019dkm,LykkeJacobsen:2018cbt}.

\subsection{Galois shuffle and monodromy charge}

In this subsection, we introduce the monodromy charge of the operators in CFTs which plays a central role in studying modular $S$ property of the cylinder partition function\cite{Gannon:2003de,Harvey:2019qzs}. The monodromy charge $Q^{\text{mon}}_{J}(\alpha)$ of a quantum state labeled by $\alpha$ under the symmetry operator $J$ is defined as,

\begin{equation}
S_{\alpha', \alpha} =e^{2i\pi Q^{\text{mon}}_{J}(\alpha)}S_{J\alpha' , \alpha}
\end{equation}
where we have used matrix representation of the modular $S$ matrix and $\alpha$ and $\alpha'$ are the label of primary fields. 

It is known that in a unitary minimal model, this monodromy charge matches with the simple charge of the same operator. In a class of nonunitary CFT, these two charges do not match and it can be represented as,
\begin{equation}
Q^{\text{mon}}_{J}(\alpha)=Q_{J}(\alpha)-Q_{J}(\sigma)\neq Q_{J}(\alpha),
\end{equation}
where $\sigma$ is the primary field with the lowest conformal dimension in the theory satisfying $S_{\sigma,\alpha}>0$ for all $\alpha$. This mismatch of the monodromy charge and the simple charge is called Galois shuffle\cite{Gannon:2003de}, and plays a central role in considering the modular $S$ property of the cylinder partition function of the FQH system. 

 Hence, when considering FQHE which couples Hilbert space by matching simple charge, one can expect the modular $S$ property of the original theory can survive under coupling. However, in nonunitary theories, such as the $M(p,q)$ minimal model outside of $q=p+1$, these two charges do not match. Hence, difficulties obtaining modular $S$ property for the FQHE system arise as we show in the following sections. Even in this case, one can construct a modular $S$ invariant of FQHE like system by adding fractional flux quantum \cite{Milovanovic:1996nj,Ino:1998by} as we discuss in the appendix, but one should keep in mind that this is different from the original FQHE. 

\section{Fermionic simple current in Gaffnian CFT}
\label{fermion}

As was noticed in older literature, $M(p,p')$ minimal model has $Z_{2}$ simple current field $\phi_{1,p'-1}=\phi_{p-1,1}$. For $M(3,5)$, $\phi(1,4)=\phi(2,1)$ with conformal dimension $\frac{3}{4}$ is the $Z_{2}$ simple current.
The relevant fusion rules are the following,
\begin{align} 
J\times\sigma&=\sigma', \\
J\times\sigma'&=\sigma, \\
\sigma \times \sigma'&=\sigma+J, \\
\sigma \times \sigma&=I+\sigma', \\
\sigma' \times \sigma'&=I+\sigma'.
\end{align}
where $\sigma=\phi_{1,2}=\phi_{2,3}$ with conformal dimension $-1/20$ and $\sigma'=\phi_{1,3}=\phi_{2,2}$ with conformal dimension $1/5$. As can be seen in the fusion rule, $\sigma$ and $\sigma'$ can have zero modes by assigning $J$ as parity odd. To avoid zero modes, one has to take $\sigma$ as odd and $\sigma'$ as even. Because the existence of the zero modes, i.e.  $\sigma=(\sigma^{\text{even}}+\sigma^{\text{odd}})/\sqrt{2}$, $\sigma'=(\sigma^{'\text{even}}+\sigma^{'\text{odd}})/\sqrt{2}$, leads to noninteger fusion rule in this semion basis, we show only the results for the case without zero modes (However, one can easily obtain the corresponding partition functions assuming the existence of the zero modes).

The chiral partition functions are,

\begin{align}
\Xi_{0,\frac{r}{q}}^{\text{even}}&= \chi_{0}\theta^{\text{even}}_{\frac{r}{q}} ,\\
\Xi_{0,\frac{r}{q}}^{\text{odd}}&= \chi_{0}\theta^{\text{even}}_{\frac{r}{q}}, \\
\Xi_{J, \frac{r}{q}}^{\text{tw, even}}&=\chi_{J}\theta^{\text{odd}}_{\frac{r+q}{q}} ,\\
\Xi_{J, \frac{r}{q}}^{\text{tw, odd}}&= \chi_{J}\theta^{\text{even}}_{\frac{r+q}{q}},
\end{align}

\begin{align}
\Xi_{\sigma,\frac{r}{q}}^{\text{even}}&=\chi_{\sigma}\theta^{\text{odd}}_{\frac{r}{q}},\\
\Xi_{\sigma,\frac{r}{q}}^{\text{odd}}&= \chi_{\sigma}\theta^{\text{even}}_{\frac{r}{q}}, \\
\Xi_{\sigma', \frac{r}{q}}^{\text{tw, even}}&=\chi_{\sigma'}\theta^{\text{odd}}_{\frac{r+q}{q}},\\
\Xi_{\sigma', \frac{r}{q}}^{\text{tw, odd}}&= \chi_{\sigma'}\theta^{\text{even}}_{\frac{r+q}{q}},
\end{align}
It should be noted that the simple current operator itself is twisted and couples to half-flux quantum and this is distinct from the anomaly free cases. For the later discussion, we introduce the character $\Xi '$ which is labeled by the same indices $\Xi$, but attached with half flux quantum involving the transformation $r\rightarrow r+1/2$. In short, one can define $\Xi'_{\frac{r}{q}}=\Xi_{\frac{r+1/2}{q}}$ where we have dropped off other indices coming from the primary fields and the parity. This $\Xi'$ does not appear from the construction based on the bulk-edge correspondence in FQHE but plays a role in considering the modular property of the system as we will see.

To construct the partition function, let us introduce a simple assumption (However, in the next section, we will see this cannot be applied to realistic situations). We assume the theory is described fermionically (or bosonically), i.e. the system is invariant under $T^{2}$ (or $T$).

Then the partition function with total fermion number even is,
\begin{equation}
\begin{split}
&Z^{\text{even}}_{T-\text{even}} \\
&=\sum_{r}\Xi_{0,\frac{r}{q}}^{\text{even}}\overline{\Xi}_{0,\frac{r}{q}}^{\text{even}}+\Xi_{0,\frac{r}{q}}^{\text{odd}}\overline{\Xi}_{0,\frac{r}{q}}^{\text{odd}}+\Xi_{J,\frac{r}{q}}^{\text{even}}\overline{\Xi}_{J,\frac{r}{q}}^{\text{even}}+\Xi_{J,\frac{r}{q}}^{\text{odd}}\overline{\Xi}_{J,\frac{r}{q}}^{\text{odd}} \\
&+\Xi_{\sigma,\frac{r}{q}}^{\text{even}}\overline{\Xi}_{\sigma,\frac{r}{q}}^{\text{even}}+\Xi_{\sigma,\frac{r}{q}}^{\text{odd}}\overline{\Xi}_{\sigma,\frac{r}{q}}^{\text{odd}}+\Xi_{\sigma',\frac{r}{q}}^{\text{even}}\overline{\Xi}_{\sigma',\frac{r}{q}}^{\text{even}}+\Xi_{\sigma',\frac{r}{q}}^{\text{odd}}\overline{\Xi}_{\sigma',\frac{r}{q}}^{\text{odd}}.
\end{split}
\end{equation}
where we have taken the summation $r=0,...,2q$, and we drop off this summation indices to avoid complications of indices in the following discussion.

Similarly, we can obtain the odd partition function by changing the label even or odd of holomorphic characters. It should be noted that these partition functions satisfy modular $T$ invariance. The total partition function is,
\begin{equation}
Z_{T-\text{inv}}=\sum_{r}\Xi_{0,\frac{r}{q}}\overline{\Xi}_{0,\frac{r}{q}}+\Xi_{J,\frac{r}{q}}\overline{\Xi}_{J,\frac{r}{q}}+\Xi_{\sigma,\frac{r}{q}}\overline{\Xi}_{\sigma,\frac{r}{q}}+\Xi_{\sigma',\frac{r}{q}}\overline{\Xi}_{\sigma',\frac{r}{q}}.
\end{equation}
where we have introduce the character $\Xi =\Xi^{\text{even}}+\Xi^{\text{odd}}$ for the each label of the fields.

\subsection{Comparison with other literature and imaginary gapping argument}

It should be noted our partition function for the even sector is consistent with the work by Jolicoeur, Mizuaki, and Lecheminant by fixing the total number of fermion or by restricting the sector of $U(1)$ \cite{Jolicoeur:2014isa}. For the odd sector, our fermionic partition function is different from the results of them. Surprisingly, it contains couplings of twisted and untwisted sectors.

Their partition function in the odd sector seems to be,

\begin{equation}
\begin{split}
&Z^{\text{odd}}_{\text{JML}} \\
&=\sum_{r}\Xi_{0,\frac{r}{q}}\overline{\Xi}_{J,\frac{r}{q}}+\Xi_{J,\frac{r}{q}}\overline{\Xi}_{0,\frac{r}{q}} \\
&+\Xi_{\sigma,\frac{r}{q}}\overline{\Xi}_{\sigma',\frac{r}{q}}+\Xi_{\sigma',\frac{r}{q}}\overline{\Xi}_{\sigma,\frac{r}{q}}
\end{split}
\end{equation}

Remarkably, these partition functions only satisfy the modular $T^{2}$ invariance and the flux sector of the chiral and antichiral modes are mixed in a nontrivial way. This is a consequence of the quantum anomaly (or twist) of $Z_{2}$ simple current\cite{Furuya:2015coa} and the similar unusual partition function has been proposed in . It is possible to interpret this odd partition function as the one constructed from our even partition function by adding $J$ and a half-flux which corresponds to a fermionic particle. The half-flux quantum breaks the modular $T^{2}$ invariance of the system and this fermionic parity shift operation can be considered as a generalized version of parity shift operation in \cite{Runkel:2020zgg,Makabe:2017ygy}. Let us focus on the structure of total partition function $Z_{T-\text{inv}}+Z^{\text{odd}}_{\text{JML}}$ in the following discussion.

It is worth noting that one has to assume fermionic parity of $M(3,5)$ to obtain this partition function started from $Z^{\text{even}}_{T\text{inv}}$. However, this type of object is difficult to represent by considering the fermionization of $1+1$ dimensional quantum lattice models because of the spin-statistics and anomalous conformal dimension of $Z_{2}$ simple current. In anomalous cases, the cylinder partition function cannot be described by the existing Lagrangian CFT because this breaks modular $T$ and $T^{2}$ invariance even in the unitary CFT. This observation is consistent with the recent proposal of nonlocal representation of anomalous $Z_{2}$ symmetry\cite{Hao:2022kxo,Chatterjee:2022tyg,Chatterjee:2022kxb}. Related discussion of spin statistics in condensed matter systems can be seen in \cite{Gaiotto:2015zta}.

Next, we discuss the attachment of half-flux quantum to the total partition function $Z_{T-\text{inv}}+Z^{\text{odd}}_{\text{JML}}$. This procedure is a generalization of $\nu=5/2$ enigma for Haldane-Rezayi states\cite{Ino:1998by}. The partition function can be easily obtained by replacing all $\Xi$ to $\Xi'$, and we represent the resultant partition function as $Z'$. As we will show in the appendix, this partition function satisfies the modular $S$ invariance. 
The coupling between $U(1)$ part and CFT part in modular $S$ invariant is different from that of FQHE in the nonunitary case. This is a consequence of the Galois shuffle in nonunitary $Z_{2}$ models. 

When one applies $Z_{2}$ operation to the partition functions, the monodromy charges are the fundamental objects. However, in considering FQHEs, the simple charges are the fundamental objects which determine half-flux quantum structures. In unitary theories, these two charges coincide, but in nonunitary CFT, this is not the case. Hence, by applying an additional half-flux quantum to Gaffnian FQHE, one can obtain a modular $S$ invariant. This is an extended version of $\nu=\frac{5}{2}$ enigma in \cite{Ino:1998by,Milovanovic:1996nj}, by considering Galois shuffle.

By (imaginary) gapping out $M(3,5)$ part under breaking bulk-edge correspondence, one can obtain free bosonic partition function as modular invariant\cite{Fukusumi_2022_f}. However, by gapping out the half-flux or integer-flux sectors of Luttinger parts as in\cite{Cappelli:2010jv}, it is impossible to obtain a modular invariant of $M(3,5)$ model because of the mixing of integer flux and half-integer sector of each edge in the total partition function. Hence, there exists a difficulty to obtain $M(3,5)$ as protected edge modes of the gapped phase. In other words, if one assumes bulk-edge correspondence for $M(3,5)$ part, it becomes difficult to obtain a gapped system. Similar phenomena are present in the Haldane-Rezayi states, and it may be natural to relate modular noninvariance to the gaplessness of the states.

On the other hand, there exist several works which have proposed Gaffnian FQHE on a cylinder can be gapped\cite{Weerasinghe_2014,Papi__2014}. There exist at least three possible scenarios to resolve the inconsistency between other works and the above analysis in this section. The first is that their Gaffnian may be bosonic trivial Gaffnian, which only couples modular invariant of $M(3,5)$ and Luttinger liquid. In $M(3,5)$ CFT, there exists no $Z_{2}$ invariant state. Hence in this system, one can consider modular $T$ and $S$ invariant and parity of the systems consistently. The second is that they have accomplished tuning gapping of $M(3,5)$ part. In our argument, we do not assume an imaginary gapping operation to maintain bulk-edge correspondence, but it may be possible by choosing an appropriate interaction. The third is that their gapping operation is breaking bulk-edge correspondence.

In summary, we have established difficulty to gap out the Gaffnian state under bulk-edge correspondence. In other words, one can say that bulk gaplessness is protected by bulk-edge correspondence.

\section{Zero modes in Haldane-Rezayi state and $331$ state}
\label{HR}
In this section, we review the correspondence between $331$ states and Haldane-Rezayi states \cite{Ino:1998by,Milovanovic:1996nj} from a modern view of fermionization.

The $331$ state is known to be constructed by Weyl fermion and chiral boson. The Weyl fermion part can be constructed from two Majorana fermions.
The primary fields of Weyl fermion have the conformal dimensions $h_{0}=0$, $h_{\psi}=\frac{1}{2}$, $h_{\sigma}=\frac{1}{8}$, $h_{\sigma}=\frac{9}{8}$. The operator $\psi$ is the $Z_{2}$ simple current. As one can see easily, $h_{\sigma}=2\times\frac{1}{16}$ and it should be interpreted as zero modes.

Remarkably the Weyl fermion has (at least) the following three modular invariants\cite{Creutzig:2013hma,Guruswamy:1996rk},
\begin{align}
Z_{\text{boson}}&=\chi_{0}\overline{\chi}_{0}+\chi_{\psi}\overline{\chi}_{\psi}+\chi_{\sigma}\overline{\chi}_{\sigma}+\chi_{\sigma'}\overline{\chi}_{\sigma'}, \\
Z_{\text{fermion}}&=|\chi_{0}+\chi_{\psi}|^{2}, \\
Z_{\text{boson'}}&=\chi_{0}\overline{\chi}_{0}+\chi_{\psi}\overline{\chi}_{\psi}+2|\chi_{\sigma}+\chi_{\sigma'}|^{2}, \\
\end{align}
Each partition function corresponds to the charge conjugated partition function, fermionic partition function, and the $Z_{2}$ orbifold respectively. The zero modes of the fields can be seen in the partition function of the orbifold. This example shows that to decide whether the $Z_{2}$ noninvariant fields have zero modes or not, it may be necessary to consider the fusion rule of primary fields and the orbifolding of the theories. For a $Z_{2}$ noninvariant field $\phi_{a}$, the fusion rule of $\phi_{a}\times J\phi_{a}$, (or $\phi_{a}\times \phi_{a}$, and $J\phi_{a}\times J\phi_{a}$) has significant information because if these fields do not have zero modes, the result of fusion rule should produce only odd (or even) parity fields. So in unitary theory, by assuming the even parity of identity fields and odd parity of simple current, it may be possible to decide whether the fields have zero modes or not by considering these fusion rules recursively.

By applying the general analysis of fermionic FQHE, there exists fermionic TQFT corresponding to the partition function $Z_{\text{fermion}}$. Hence there exist at least two procedures to obtain modular $S$ invariant as in the Pfaffian case with total parity even partition function in \cite{Milovanovic:1996nj,Ino:1998by} and product of fermionic partition $Z_{\text{fermion}}$ and Luttinger liquid\cite{Cappelli:2010jv}.

As is known in the existing literature, there exists a duality between the Weyl fermion and the symplectic fermion\cite{Guruswamy:1996rk}. The most notable thing in this duality is the twisted (untwisted) sector in Weyl fermion goes to the untwisted (twisted) sector in the symplectic fermion when considering the simple charge. Hence by attaching the half flux quantum, the behavior under modular $S$ transformation changes drastically. Therefore in the Haldane-Rezayi states, it is difficult to obtain the modular invariant partition function by the imaginary gapping operation for half flux quantum sectors. This is the same as in the Gaffnian states by considering the Galois shuffle of $M(3, 5)$, as we have discussed in the previous section.

\section{conclusion}
\label{conclusion}

\begin{figure}[htbp]
\begin{center}
\includegraphics[width=0.5\textwidth]{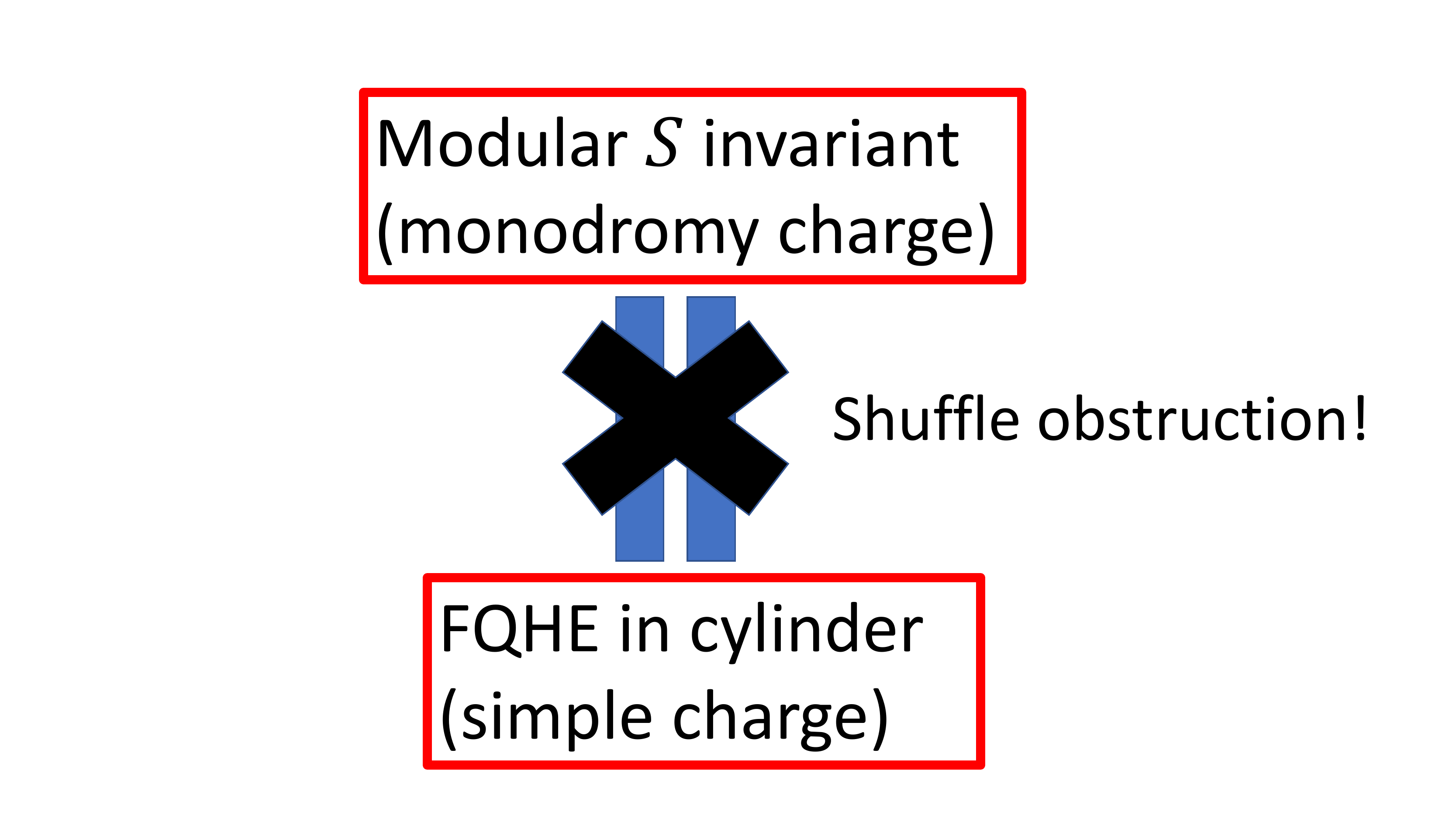}
\caption{Galois shuffle obstraction in FQHE}\label{shuffle}
\end{center}
\end{figure}

In this work, assuming modular S invariance for the cylinder partition function as a necessary condition to gap out the bulk\cite{2013PhRvX...3b1009L}, we have shown difficulty to obtain gapped bulk under bulk edge correspondence in the two classes of the models, Haldane-Rezayi and Gaffnian FQHE. Moreover, we have shown this difficulty coming from the (Galois) shuffle structure of the models (Fig. \ref{shuffle}). Hence applying the analysis of our previous work \cite{Fukusumi_2022_f} and this work, one can obtain similar difficulties systematically. It may be interesting to apply our analysis to the existing gapless SPT and analyze bulk and boundary RG under bulk irrelevant perturbation.

Whereas the two models, the Haldane-Rezayi model, and Gaffnian FQHE, seem gapless, the modular $T$ properties of these models seem different. In the Haldane-Rezayi model, the cylinder partition function satisfies the modular $T^{2}$ invariance and it seems "local" to some extent. However, the cylinder partition function of Gaffnian FQHE does not satisfy modular $T^{2}$ invariance because of the anomalous conformal dimension of the $Z_{2}$ simple current. Moreover, this anomalous conformal dimension is outside of the spin statistics theorem. Hence there exists a difficulty to represent this partition function by the existing $1+1$ dimensional fermionic lattice models.\footnote{There has existed the similar problem when mapping from thin torus limit of FQHE to $1+1$ dimensional lattice models\cite{Nakamura_2012,https://doi.org/10.48550/arxiv.1204.5682,Papi__2014}.} The similar nonlocality coming from quantum anomaly has been proposed in the parafermionic chain, but $Z_{2}$ is special because the anomalous quantum dimension breaks the spin statistics. We expect that these types of nonlocality coming from the quantum anomaly and breaking spin statistics under bulk-edge correspondence are universal phenomena. Hence, it may be natural to consider that a part of gapless topological order and deconfined quantum criticality can be considered as a consequence of gauging anomalous finite group symmetry\cite{Tachikawa:2017gyf,Bhardwaj:2017xup}.

\section{Acknowledgement}
The author thanks Yunqin Zheng and Yuji Tachikawa for the fruitful collaboration closely related to this project and also thanks Yang Bo for the other related collaboration. He also thanks helpful comments and the discussions with Yang Bo and Yuji Tachikawa. He also thanks Masahiko G. Yamada for the discussion. A large part of this work was done in the conference "Topological interacting electron in person" at Quy Nhon, Vietnam.

\appendix

\section{Nonuniqueness of Jordan-Wigner transformations}

It might be surprising but there exist two distinct Jordan-Wigner transformations and this results in two different fermionic chains which have different conformal spin when the $Z_{2}$ simple current is anomalous.

In this section, we take Pauli matrices which are usually used in the analysis of spin chains, and follow the analysis in \cite{Yao:2020dqx}. For example, in Ising model, the Hamiltonian is $H_{spin}=-\sum_{i}\left( \sigma^{z}_{i}\sigma^{z}_{i+1}+\sigma^{x}_{i}\right)$.

One Jordan-Wigner transformation is defined as 
\begin{equation}
\gamma_{2j-1}=\sigma_{j}^{z}\prod_{i=1}^{j-1}\sigma_{i}^{x}, \gamma_{2j}=i\sigma_{j}^{z}\prod_{i=1}^{j}\sigma^{x}_{i}
\end{equation}

there exists the other distinct Jordan-Wigner transformation,

\begin{equation}
\gamma_{2j-1}'=\sigma_{j}^{z}\prod_{i=j}^{L}\sigma_{i}^{x}, \gamma_{2j}'=i\sigma^{z}_{j}\prod_{i=j+1}^{L}\sigma^{x}_{i}
\end{equation}
These two Jordan-Wigner transformations are related to each other by the nonlocal unitary transformation and space inversion symmetry $\tau \leftrightarrow \overline{\tau}$. But they are living in different Hilbert spaces intrinsically. 
For example, when one acts unitary transformation $\sigma^{y}$ for a cite, these two fermion acts differently. Moreover, the action $\tau \leftrightarrow \overline{\tau}$ is not a conformal transformation because it exchanges the holomorphic part and antiholomorphic part of the theory.

There exists a bosonic $Z_{2}$ symmetry operator which appears as topological defect (or symmetry defect),
\begin{equation}
Q_{b}=\prod_{i}\sigma^{x}_{i}
\end{equation}

There also exist other fermionic symmetry operators,

\begin{align}
Q_{f}&=\prod_{j}i\left( -i \gamma_{2j-1}\gamma_{2j}\right), \\
Q_{f'}&=\prod_{j}i\left( -i \gamma_{2j-1}'\gamma'_{2j}\right)
\end{align}

The remarkable feature of these operations are
\begin{equation}
Q_{f}Q_{f'}=1
\label{twist-duality}
\end{equation}

Hence if one applies both fermionic $Z_{2}$ operations to bosonic Hilbert space, the resultant Hilbert space is invariant under this operation. 

The chiral $Z_{2}$ twist and the antichiral $Z_{2}$ twist in the field theory can be interpreted as field theoretic analog of above $Q_{f}$ and $Q_{f'}$. The chiral $Z_{2}$ twist is a line object which acts like $Z_{2}$ Verlinde line for chiral Hilbert space whereas it acts trivially on antichiral Hilbert space.

\begin{figure}[htbp]
\begin{center}
\includegraphics[width=0.5\textwidth]{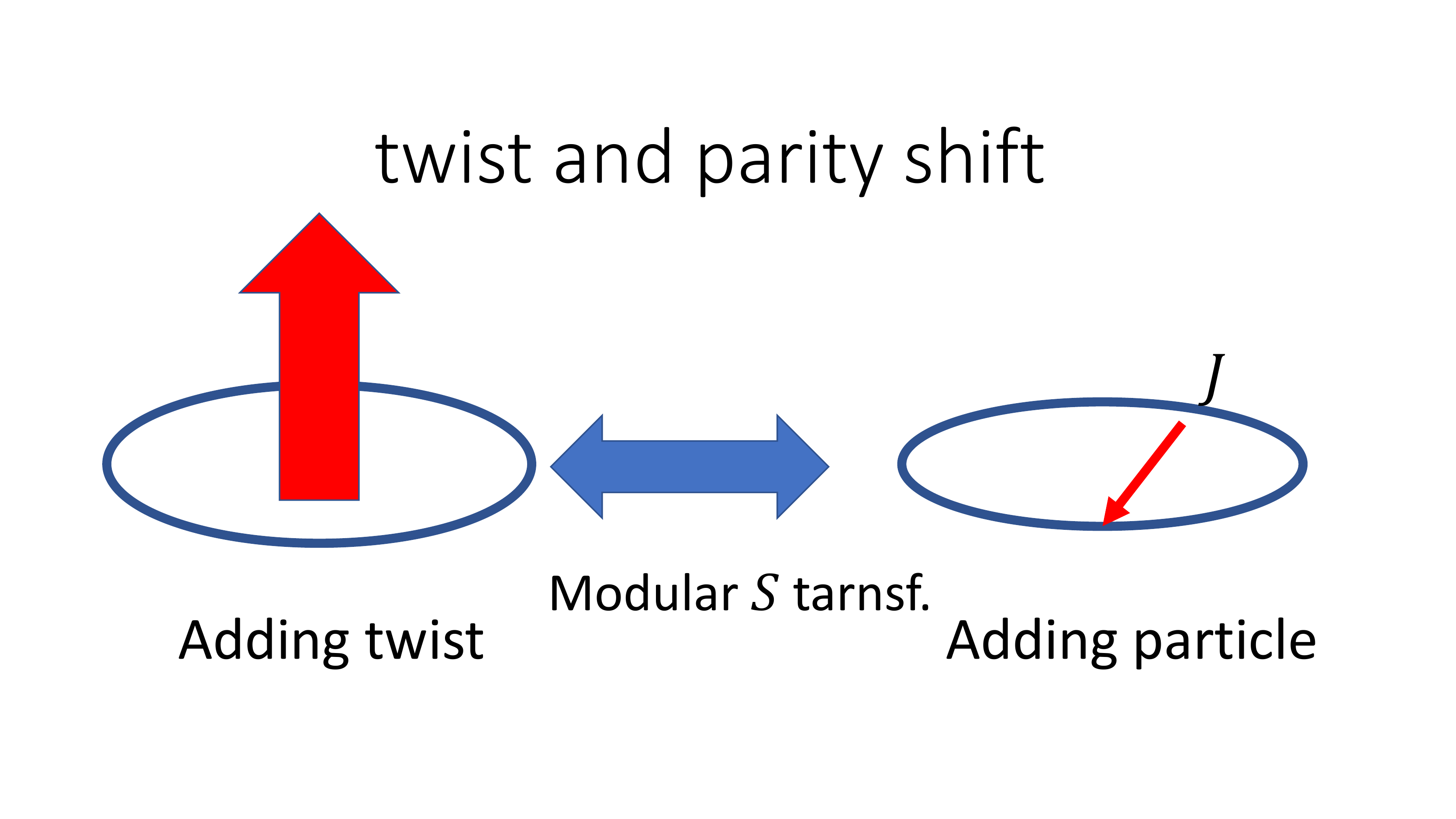}
\caption{Equivalence between twist and parity shift}\label{twist_parity}
\end{center}
\end{figure}

The remarkable feature of bosonic Hilbert space constructed from the charge conjugated modular invariant partition function is the Hilbert space can be represented as the direct summations of the tensor product of charge conjugated chiral and antichiral parts. This is a tautology, but this is essentially important when considering these $Z_{2}$ twists. Because of this charge conjugate property of bosonic Hilbert space, when acting both chiral $Z_{2}$ twist and antichiral $Z_{2}$ twist, the total bosonic Hilbert space is invariant. This is in analogy with \eqref{twist-duality}. Moreover, by interpreting modular $S$ transformation to this chiral and antichiral twist, one can obtain the chiral parity shift operation and antichiral parity shift operation which are proposed in the work by Runkel and Watts\cite{Runkel:2020zgg,Makabe:2017ygy}. Schematically, one can understand this modular $S$ property as an equivalence between twisting and adding a fermionic particle (FIG. \ref{twist_parity}).

\begin{figure}[htbp]
\begin{center}
\includegraphics[width=0.5\textwidth]{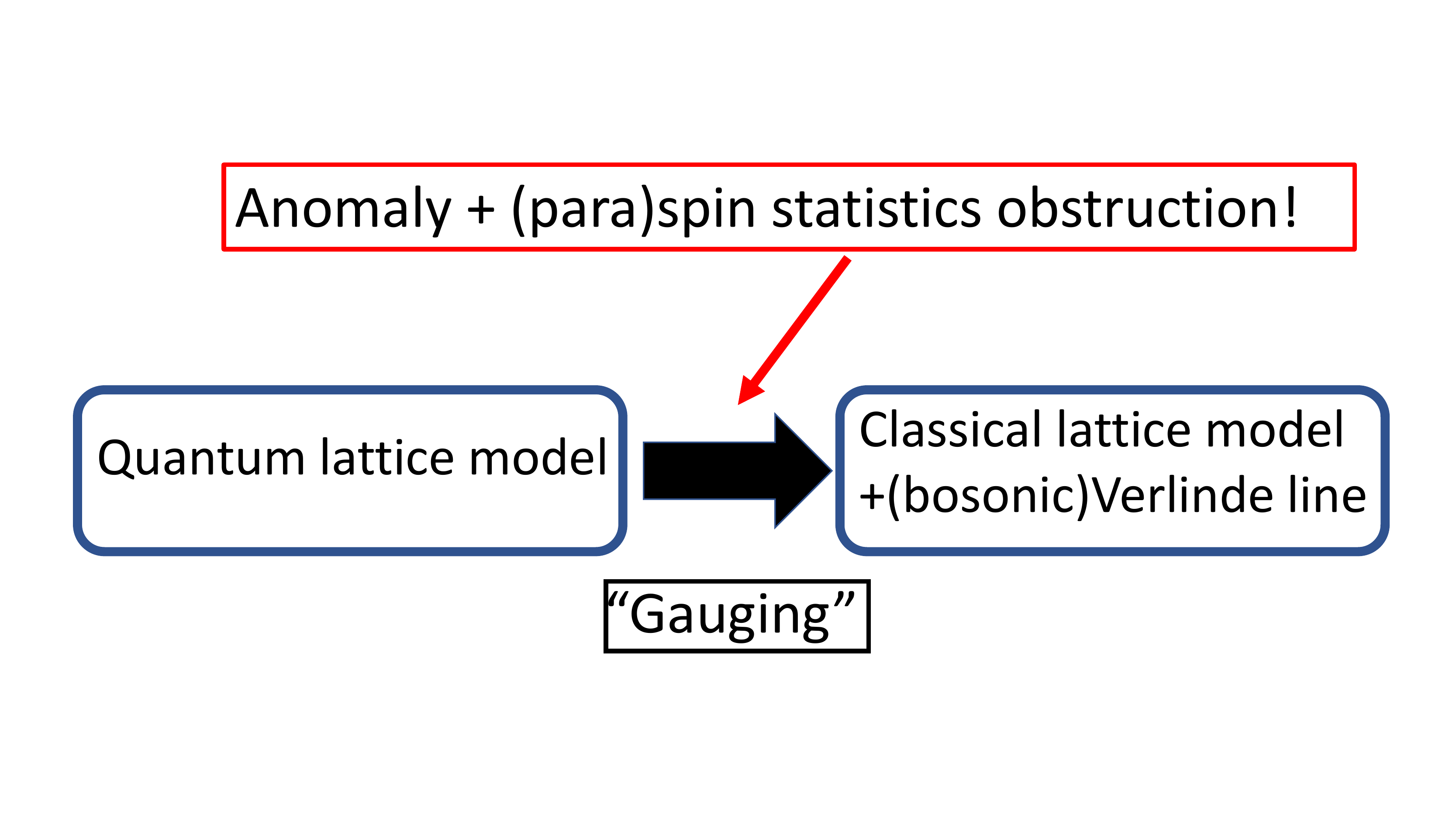}
\caption{Anomaly and (para)spin statistics obstruction in quantum-classical correspondence}\label{anomaly_spin}
\end{center}
\end{figure}

It should be remarkable that in general, there exist more topological defects in fermionic field theory whereas there exists Jordan-Wigner duality of the theories\cite{Makabe:2017ygy,Chang:2022hud}. Considering its action to Hilbert space, there exist fermionic topological defects which cannot be represented as bosonic topological defects called "Verlinde line"\cite{Petkova:2000ip}. However, the fermionic partition functions can be reproduced from the bosonic partition functions in the nonanomalous case. This is not true for the anomalous case. By interpreting the bosonic partition functions as that of 2d statistical models, this inconsistency can be interpreted as a collapse of correspondence between 1d quantum systems and 2d statistical mechanical systems. Hence existing Lagrangian expression (this is classical to some extent) constructed by $Z_{2}$ Verlinde line or $Z_{2}$ gauging cannot give anomalous fermionizations which correspond to quantum fermionic systems(FIG. \ref{anomaly_spin}). To overcome this difficulty, it is necessary to implement these chiral or antichiral $Z_{2}$ gauging in the Lagrangian formalism\footnote{Related discussion about chiral fermion in Lagrangian formalism can be seen several works, for example, see \cite{Faddeev:1984jp,Hosono:1987na,Fujiwara:1988zc}}.

The specialty of $Z_{2}$ simple current is that the condition which becomes anomalous automatically breaks the fermionic spin statistics, $h_{J}=\text{integer}, \text{half-integer}$. Hence there exists a difficulty to map the cylinder partition function of $Z_{2}$ anomalous topologically ordered states to the existing Majorana fermionic spin chains. All the above arguments can be naturally generalized to $Z_{N}$ symmetric case, as the author will discuss in the other related work \cite{Fukusumi_2022_c}.

\subsection{Application of the $Z_{2}$ chiral and antichiral twists to $SU(2)_{1}$}

In this section, we introduce the $Z_{2}$ chiral and antichiral transformations to $SU(2)_{1}$ model, as the simplest model. We define the action of the chiral $Z_{2}$ twist $\mathcal{L}_{f}$ to the Hilbert space of $SU(2)_{1}$ as the following,

\begin{align} 
\mathcal{L}_{f}|0\rangle \overline{|0\rangle}&=|0\rangle \overline{|0\rangle}, \\
\mathcal{L}_{f}|0\rangle \overline{|\phi_{\frac{1}{2}}\rangle}&=|0\rangle \overline{|\phi_{\frac{1}{2}}\rangle}, \\
\mathcal{L}_{f}|\phi_{\frac{1}{2}}\rangle \overline{|0\rangle}&=-|\phi_{\frac{1}{2}}\rangle \overline{|0\rangle}, \\
\mathcal{L}_{f}|\phi_{\frac{1}{2}}\rangle \overline{|\phi_{\frac{1}{2}}\rangle}&=-|\phi_{\frac{1}{2}}\rangle \overline{|\phi_{\frac{1}{2}}\rangle}, \\
\end{align}

where $0$ and $\phi_{\frac{1}{2}}$ are the identity operator and the spin $\frac{1}{2}$ primary field respectively. By applying modular $S$ transformation to this operation, the chiral twist becomes chiral parity shift $\mathcal{L}_{p}$ which acts on chiral Hilbert space, 
\begin{align} 
\mathcal{L}_{p}|0\rangle \overline{|0\rangle}&=|\phi_{\frac{1}{2}}\rangle \overline{|0\rangle}, \\
\mathcal{L}_{p}|0\rangle \overline{|\phi_{\frac{1}{2}}\rangle}&=|\phi_{\frac{1}{2}}\rangle \overline{|\phi_{\frac{1}{2}}\rangle}, \\
\mathcal{L}_{p}|\phi_{\frac{1}{2}}\rangle \overline{|0\rangle}&=|0\rangle \overline{|0\rangle}, \\
\mathcal{L}_{p}|\phi_{\frac{1}{2}}\rangle \overline{|\phi_{\frac{1}{2}}\rangle}&=|0\rangle \overline{|\phi_{\frac{1}{2}}\rangle},
\end{align}

Hence by applying this twist to space and time directions respectively, we can obtain the partition functions as,
\begin{align}
Z_{A}&=|\chi_{0}|^{2}+|\chi_{\frac{1}{2}}|^{2}, \\
Z_{t}&=|\chi_{0}|^{2}-|\chi_{\frac{1}{2}}|^{2}, \\
Z_{s}&=\chi_{0}\overline{\chi_{\frac{1}{2}}}+\chi_{\frac{1}{2}}\overline{\chi_{0}},\\
Z_{st}&=\chi_{0}\overline{\chi_{\frac{1}{2}}}-\chi_{\frac{1}{2}}\overline{\chi_{0}},
\end{align}
where $Z_{A}$ is the charge conjugated modular invariant and the lower index $s$, $t$ labels the insertion of twists.
In the same manner, one can obtain the set of partition functions by acting antichiral twists $\mathcal{L}_{f'}$, 

\begin{align}
Z_{A}&=|\chi_{0}|^{2}+|\chi_{\frac{1}{2}}|^{2}, \\
Z_{t}'&=|\chi_{0}|^{2}-|\chi_{\frac{1}{2}}|^{2}, \\
Z_{s}'&=\chi_{0}\overline{\chi_{\frac{1}{2}}}+\chi_{\frac{1}{2}}\overline{\chi_{0}},\\
Z_{st}'&=-\chi_{0}\overline{\chi_{\frac{1}{2}}}+\chi_{\frac{1}{2}}\overline{\chi_{0}},
\end{align}

\begin{figure}[htbp]
\begin{center}
\includegraphics[width=0.5\textwidth]{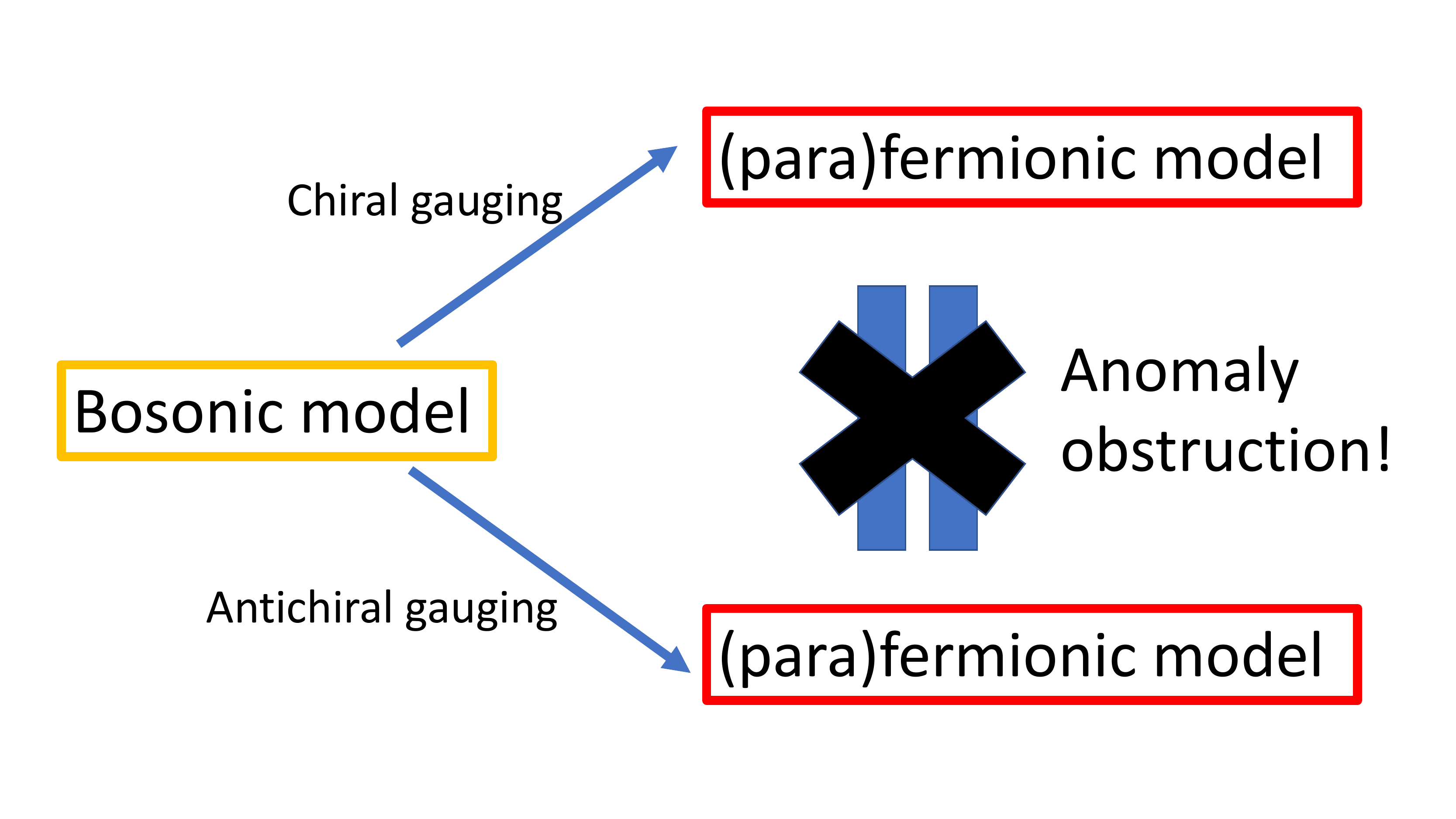}
\caption{Anomaly obstraction and emergence of chirality}\label{anomaly_chirality}
\end{center}
\end{figure}

Because $Z_{st}$ and $Z_{st}'$ are different, the resultant fermionic CFTs have different structures (FIG.\ref{anomaly_chirality}).

\subsection{Anomalous simple charge of simple current and anomalous fermionization}

In this section, we summarize a general procedure to obtain modular $S$ invariant partition functions starting from a modular invariant partition function by considering the chiral twist and chiral parity shift.

First, let us assume the existence of the modular invariant $Z_{A}$ and simple current $J$. In this setting, one can consider the parity shift of the chiral part, as $\chi_{a}\rightarrow \chi_{Ja}$. There exist two directions, space and time, to realize this operation in the torus geometry. In the spin chain, the insertion of this operation in the space direction can be done by implementing the corresponding defect, and that in the time direction can be done by implementing the equation of motion (or topological symmetry operator in modern literature). It should be noted that the resultant partition functions by these operations, $Z_{s}$ and $Z_{t}$ respectively, are related to each other by the modular $S$ transformation. Hence it is possible to obtain the modular invariant as
\begin{equation}
 Z_{s}+Z_{t}.
\end{equation}
However, it should be noted that the above partition function does not become a realistic partition function because the coefficients can become outside of the positive integer.  

By considering the parity shift operation both in the time direction and space direction, one can obtain another partition function $Z_{st}$. If the simple current operator is nonanomalous, this becomes a modular $S$ invariant. However, if the simple current is anomalous, the modular $S$ transformation for $Z_{st}$ changes that are constructed from the antichiral parity shift. Hence one can implement modular covariants as,

\begin{equation}
S: \ Z_{st}\rightarrow\overline{Z_{st}}
\end{equation}

Let us assume the anomalous bosonic theory is described by the following charge conjugated partition functions,

\begin{equation} 
Z_{A}=\sum_{i}|\chi_{i}|^{2}+|\chi_{i_{\text{tw}}}|^{2}
\end{equation}
where $i$ is labeling the untwisted sector with $Q_{J}(i)=0$ and $i_{\text{tw}}$ is labeling the field twisted by anomalous simple current $J$ to the field labeled by $i$.

When the theory is unitary, the fermionic partition function produced by the chiral twist corresponding to Neveu-Schwartz (NS) sector is,
\begin{equation}
Z_{f,NS}=\sum_{i}|\chi_{i}|^{2}+\chi_{i}\overline{\chi}_{i_{\text{tw}}},
\end{equation}
where we have used the notation that the simple charge of the chiral sector becomes $0$. One can obtain the partition function corresponding to the Ramond sector in the same way as,
\begin{equation}
Z_{f,R}=\sum_{i}|\chi_{i_{\text{tw}}}|^{2}+\chi_{i_{\text{tw}}}\overline{\chi_{i}},
\end{equation}

In the nonunitary anomalous case, the fermionic partition functions are,
\begin{equation}
Z_{f,NS}=\sum_{i}|\chi_{i_{\text{tw}}} |^{2}+\chi_{i_{\text{tw}}}\overline{\chi_{i}},
\end{equation}
\begin{equation}
Z_{f,R}=\sum_{i}|\chi_{i}|^{2}+\chi_{i}\overline{\chi_{i_{\text{tw}}}},
\end{equation}
because of the Galois shuffle, one can see the simple charges of $NS$ and $R$ sector is shuffled compared with the unitary case.

Hence by considering the summation of these modular $S$ invariants and covariants, one can construct modular $S$ invariant partition functions. The equivalent modular $S$ invariant can be obtained by directly applying $Z_{2}$ Verlinde line as in the manner of orbifolding. 

When considering FQHE systems, there exists the other $Z_{2}$ transformation, the insertion of half-flux quantum. The chiral half flux attachment changes the bosonic partition functions as follows,

\begin{align}  
Z_{L}&=\sum_{r}\theta^{+}_{\frac{r}{q}}\overline{\theta}^{+}_{\frac{r}{q}}, \\
Z_{L,t}&= \sum_{r}\theta_{\frac{r+1/2}{q}}^{+}\overline{\theta}_{\frac{r}{q}}^{+},\\
Z_{L,s}&= \sum_{r}\theta^{-}_{\frac{r}{q}}\overline{\theta}_{\frac{r}{q}}^{+},\\
Z_{L,st}&=\sum_{r}\theta^{-}_{\frac{r+1/2}{q}}\overline{\chi_{\frac{r}{q}}} ,\\
\end{align}
where $Z_{L}$ corresponds to the partition function of the Laughlin state with filling factor $1/q$ in the cylinder geometry, and we have taken the summation $r=0, 1, ...,q-1$ as in the main text.

Next, we note the modular $S$ transformation for bosonic characters from the paper by Ino\cite{Ino:1998by},
\begin{align}
\theta_{\frac{r}{q}}^{\text{even}}&\rightarrow \frac{1}{2\sqrt{q}} \sum_{s=0}^{q-1}\left(e^{\frac{2\pi irs}{q}}\theta^{+}_{\frac{s}{q}}+e^{\frac{2\pi ir\left(s+\frac{1}{2}\right)}{q}}\theta^{-}_{\frac{s}{q}}\right), \\
\theta_{\frac{r}{q}}^{\text{odd}}&\rightarrow \frac{1}{2\sqrt{q}}\sum_{s=0}^{q-1}\left( e^{\frac{2\pi irs}{q}}\theta^{+}_{\frac{s}{q}}-e^{\frac{2\pi ir\left(s+\frac{1}{2}\right)}{q}}\theta^{-}_{\frac{s}{q}}\right), \\
\theta_{\frac{r+1/2}{q}}^{\text{even}}&\rightarrow \frac{1}{2\sqrt{q}} \sum_{s=0}^{q-1}\left(e^{\frac{2\pi irs}{q}}\theta^{+}_{\frac{s}{q}}+e^{\frac{2\pi ir\left(s+\frac{1}{2}\right)}{q}}\theta^{-}_{\frac{s}{q}}\right), \\
\theta_{\frac{r+1/2}{q}}^{\text{odd}}&\rightarrow \frac{1}{2\sqrt{q}} \sum_{s=0}^{q-1}\left( e^{\frac{2\pi irs}{q}}\theta^{+}_{\frac{s}{q}}-e^{\frac{2\pi ir\left(s+\frac{1}{2}\right)}{q}}\theta^{-}_{\frac{s}{q}}\right),
\end{align}

In \cite{Ino:1998by}, the author has constructed a series of modular $S$ invariants of the FQHE model constructed from $SU(2)_{k}$ WZW model by considering the "spinon" basis for $SU(2)_{K}$ WZW model even with $K$ odd. His analysis can be easily generalized to other anomalous unitary $Z_{2}$ FQH models. Hence, we conjecture that the following partition function satisfies the modular $S$ invariant constructed by changing the basis,

\begin{equation}
Z=\sum_{r}|\Xi_{i,r}+\Xi_{i_{\text{tw}},r}|
\end{equation}
where we have assumed the FQHE construction of the character $\Xi$ as in the main text by considering the simple charge of each CFT operator.
It should be noted that when $\tau$ is pure imaginary, this exactly satisfies the modular $S$ invariance. 
For nonunitary cases, one can obtain the modular $S$ invariant in the same way, by exchanging $i$ and $i_{\text{tw}}$ or $\Xi$ to $\Xi'$ by attaching half-flux quantum \cite{Milovanovic:1996nj,Ino:1998by} as we have explained in the main text.

\section{Modular $S$ transformation for $M(3,5)$}

Here we note the modular $S$ transformation property of the $M(3,5)$ minimal model.

First, the central charge and conformal dimensions of the $M(p,p')$ minimal model are,
\begin{align}
c&=1-\frac{(p-p')^{2}}{4pp'}, \\
h_{r,s}&=\frac{(p'r-ps)^{2}-(p-p')^{2}}{4pp'}.
\end{align}
Then corresponding modular $S$ matrix is,
\begin{equation}
S_{r,s;r's'}=2\sqrt{\frac{2}{pp'}}(-1)^{1+rs'+sr'}\text{sin}\left( \frac{\pi rr'p'}{p}\right)\text{sin}\left( \frac{\pi ss'p}{p'}\right)
\end{equation}
Hence by identifying $p=3$, $p'=5$, we can obtain the desired modular matrix for the $M(3,5)$ minimal model.

There exist at least two modular invariant partition functions.
First, we note here the charge conjugated modular partition function,
\begin{equation}
Z_{A}=\sum_{i} \chi_{\alpha}\overline{\chi_{\alpha}}
\end{equation}
where $\alpha$ is labeling the primary fields of the theory.

 Next, by considering the $Z_{2}$ transformation corresponding to the twist of fields, we can obtain the following fermionic partition functions,
\begin{equation}
Z_{NS,f}=\frac{Z_{A}+Z_{s}+Z_{t}+Z_{st}}{2}=|\chi_{\sigma}|^{2}+|\chi_{J}|^{2}+\chi_{\sigma}\overline{\chi_{\sigma'}}+\chi_{J}\overline{\chi_{0}}
\end{equation}
By considering antichiral $Z_{2}$ transformation,
\begin{equation}
Z_{NS,f'}=\frac{Z_{A}+Z'_{s}+Z'_{t}+Z'_{st}}{2}=|\chi_{\sigma}|^{2}+|\chi_{J}|^{2}+\chi_{\sigma'}\overline{\chi_{\sigma}}+\chi_{0}\overline{\chi_{J}}
\end{equation}
Similar partition functions can appear by considering $Z_{3}$ model\cite{Yao:2020dqx}. It should be noted that these partition functions satisfy the following rule under modular $S$ transformation, $S: Z_{f}\leftrightarrow, Z_{f'}$. Because of the anomalous conformal dimension of the simple current, these two operations result in different partition functions.
The summation of these two partition functions is modular $S$ invariant and it corresponds to the partition function constructed from the usual $Z_{2}$ Verlinde line,
\begin{equation}
2Z_{D}=2|\chi_{\sigma}|^{2}+2|\chi_{J}|^{2}+\chi_{\sigma}\overline{\chi_{\sigma'}}+\chi_{J}\overline{\chi_{0}}+\chi_{\sigma'}\overline{\chi_{\sigma}}+\chi_{0}\overline{\chi_{J}}.
\end{equation}

As is considered in other fermionic CFTs, it may have physical meaning to consider the Hilbert space $0 \otimes J \oplus \sigma \otimes \sigma'$. Unfortunately, the corresponding partition function does not satisfy modular $S$ invariance. However, by summing the original partition function and its modular $S$ dual, one can obtain the following simple modular $S$ invariant partition function,
\begin{equation}
2Z_{D}+Z_{A}=|\chi_{0}+\chi_{J}|^{2}+|\chi_{\sigma}+\chi_{\sigma'}|^{2}+2|\chi_{J}|^{2}+2|\chi_{\sigma}|^{2}
\end{equation}

A similar partition function can be seen in \cite{LU198946}.
It should be noted that all of the above partition functions in this section, except for the usual charge conjugated partition function, do not satisfy the modular $T^{2}$ invariance. They only satisfy $T^{4}$ invariance. Moreover, they inevitably contain the mixing channel of twisted and untwisted characters. Hence, by considering the coupling with chiral Luttinger liquid, it is impossible to obtain the protected edge modes corresponding to these partition functions only by the imaginary gapping operation to half-flux quantum sectors as in \cite{Cappelli:2010jv,Fukusumi_2022_f}.

\bibliography{Gapless}

\begin{thebibliography}{70}
\expandafter\ifx\csname natexlab\endcsname\relax\def\natexlab#1{#1}\fi
\expandafter\ifx\csname bibnamefont\endcsname\relax
  \def\bibnamefont#1{#1}\fi
\expandafter\ifx\csname bibfnamefont\endcsname\relax
  \def\bibfnamefont#1{#1}\fi
\expandafter\ifx\csname citenamefont\endcsname\relax
  \def\citenamefont#1{#1}\fi
\expandafter\ifx\csname url\endcsname\relax
  \def\url#1{\texttt{#1}}\fi
\expandafter\ifx\csname urlprefix\endcsname\relax\def\urlprefix{URL }\fi
\providecommand{\bibinfo}[2]{#2}
\providecommand{\eprint}[2][]{\url{#2}}

\bibitem[{\citenamefont{Furuya and Oshikawa}(2017)}]{Furuya:2015coa}
\bibinfo{author}{\bibfnamefont{S.~C.} \bibnamefont{Furuya}} \bibnamefont{and}
  \bibinfo{author}{\bibfnamefont{M.}~\bibnamefont{Oshikawa}},
  \bibinfo{journal}{Phys. Rev. Lett.} \textbf{\bibinfo{volume}{118}},
  \bibinfo{pages}{021601} (\bibinfo{year}{2017}), \eprint{1503.07292}.

\bibitem[{\citenamefont{Numasawa and Yamaguchi}(2018)}]{Numasawa:2017crf}
\bibinfo{author}{\bibfnamefont{T.}~\bibnamefont{Numasawa}} \bibnamefont{and}
  \bibinfo{author}{\bibfnamefont{S.}~\bibnamefont{Yamaguchi}},
  \bibinfo{journal}{JHEP} \textbf{\bibinfo{volume}{11}}, \bibinfo{pages}{202}
  (\bibinfo{year}{2018}), \eprint{1712.09361}.

\bibitem[{\citenamefont{Kikuchi and Zhou}(2019)}]{Kikuchi:2019ytf}
\bibinfo{author}{\bibfnamefont{K.}~\bibnamefont{Kikuchi}} \bibnamefont{and}
  \bibinfo{author}{\bibfnamefont{Y.}~\bibnamefont{Zhou}}
  (\bibinfo{year}{2019}), \eprint{1908.02918}.

\bibitem[{\citenamefont{Han et~al.}(2017)\citenamefont{Han, Tiwari, Hsieh, and
  Ryu}}]{Han:2017hdv}
\bibinfo{author}{\bibfnamefont{B.}~\bibnamefont{Han}},
  \bibinfo{author}{\bibfnamefont{A.}~\bibnamefont{Tiwari}},
  \bibinfo{author}{\bibfnamefont{C.-T.} \bibnamefont{Hsieh}}, \bibnamefont{and}
  \bibinfo{author}{\bibfnamefont{S.}~\bibnamefont{Ryu}},
  \bibinfo{journal}{Phys. Rev. B} \textbf{\bibinfo{volume}{96}},
  \bibinfo{pages}{125105} (\bibinfo{year}{2017}), \eprint{1704.01193}.

\bibitem[{\citenamefont{Cardy}(2004)}]{Cardy:2004hm}
\bibinfo{author}{\bibfnamefont{J.~L.} \bibnamefont{Cardy}}
  (\bibinfo{year}{2004}), \eprint{hep-th/0411189}.

\bibitem[{\citenamefont{Lencses et~al.}(2019)\citenamefont{Lencses, Viti, and
  Takacs}}]{Lencses:2018paa}
\bibinfo{author}{\bibfnamefont{M.}~\bibnamefont{Lencses}},
  \bibinfo{author}{\bibfnamefont{J.}~\bibnamefont{Viti}}, \bibnamefont{and}
  \bibinfo{author}{\bibfnamefont{G.}~\bibnamefont{Takacs}},
  \bibinfo{journal}{JHEP} \textbf{\bibinfo{volume}{01}}, \bibinfo{pages}{177}
  (\bibinfo{year}{2019}), \eprint{1811.06500}.

\bibitem[{\citenamefont{Gannon}(2002)}]{Gannon:2001ki}
\bibinfo{author}{\bibfnamefont{T.}~\bibnamefont{Gannon}},
  \bibinfo{journal}{Nucl. Phys. B} \textbf{\bibinfo{volume}{627}},
  \bibinfo{pages}{506} (\bibinfo{year}{2002}), \eprint{hep-th/0106105}.

\bibitem[{\citenamefont{Yao and Furusaki}(2020)}]{Yao:2020dqx}
\bibinfo{author}{\bibfnamefont{Y.}~\bibnamefont{Yao}} \bibnamefont{and}
  \bibinfo{author}{\bibfnamefont{A.}~\bibnamefont{Furusaki}}
  (\bibinfo{year}{2020}), \eprint{2012.07529}.

\bibitem[{\citenamefont{Fendley}(2012)}]{Fendley:2012vv}
\bibinfo{author}{\bibfnamefont{P.}~\bibnamefont{Fendley}}, \bibinfo{journal}{J.
  Stat. Mech.} \textbf{\bibinfo{volume}{1211}}, \bibinfo{pages}{P11020}
  (\bibinfo{year}{2012}), \eprint{1209.0472}.

\bibitem[{\citenamefont{Fukusumi and Iino}(2020)}]{Fukusumi:2020irh}
\bibinfo{author}{\bibfnamefont{Y.}~\bibnamefont{Fukusumi}} \bibnamefont{and}
  \bibinfo{author}{\bibfnamefont{S.}~\bibnamefont{Iino}}
  (\bibinfo{year}{2020}), \eprint{2004.04415}.

\bibitem[{\citenamefont{Cappelli and Zemba}(1997)}]{Cappelli:1996np}
\bibinfo{author}{\bibfnamefont{A.}~\bibnamefont{Cappelli}} \bibnamefont{and}
  \bibinfo{author}{\bibfnamefont{G.~R.} \bibnamefont{Zemba}},
  \bibinfo{journal}{Nucl. Phys. B} \textbf{\bibinfo{volume}{490}},
  \bibinfo{pages}{595} (\bibinfo{year}{1997}), \eprint{hep-th/9605127}.

\bibitem[{\citenamefont{Cappelli et~al.}(1999)\citenamefont{Cappelli, Georgiev,
  and Todorov}}]{Cappelli:1998ma}
\bibinfo{author}{\bibfnamefont{A.}~\bibnamefont{Cappelli}},
  \bibinfo{author}{\bibfnamefont{L.~S.} \bibnamefont{Georgiev}},
  \bibnamefont{and} \bibinfo{author}{\bibfnamefont{I.~T.}
  \bibnamefont{Todorov}}, \bibinfo{journal}{Commun. Math. Phys.}
  \textbf{\bibinfo{volume}{205}}, \bibinfo{pages}{657} (\bibinfo{year}{1999}),
  \eprint{hep-th/9810105}.

\bibitem[{\citenamefont{Cappelli and Randellini}(2013)}]{Cappelli:2013iga}
\bibinfo{author}{\bibfnamefont{A.}~\bibnamefont{Cappelli}} \bibnamefont{and}
  \bibinfo{author}{\bibfnamefont{E.}~\bibnamefont{Randellini}},
  \bibinfo{journal}{JHEP} \textbf{\bibinfo{volume}{12}}, \bibinfo{pages}{101}
  (\bibinfo{year}{2013}), \eprint{1309.2155}.

\bibitem[{\citenamefont{Park et~al.}(2017)\citenamefont{Park, Fang, Bernevig,
  and Gilbert}}]{Park:2016xhc}
\bibinfo{author}{\bibfnamefont{M.~J.} \bibnamefont{Park}},
  \bibinfo{author}{\bibfnamefont{C.}~\bibnamefont{Fang}},
  \bibinfo{author}{\bibfnamefont{B.~A.} \bibnamefont{Bernevig}},
  \bibnamefont{and} \bibinfo{author}{\bibfnamefont{M.~J.}
  \bibnamefont{Gilbert}}, \bibinfo{journal}{Phys. Rev. B}
  \textbf{\bibinfo{volume}{95}}, \bibinfo{pages}{235130}
  (\bibinfo{year}{2017}), \eprint{1604.00407}.

\bibitem[{\citenamefont{Chen et~al.}(2016)\citenamefont{Chen, Tiwari, and
  Ryu}}]{Chen_2016}
\bibinfo{author}{\bibfnamefont{X.}~\bibnamefont{Chen}},
  \bibinfo{author}{\bibfnamefont{A.}~\bibnamefont{Tiwari}}, \bibnamefont{and}
  \bibinfo{author}{\bibfnamefont{S.}~\bibnamefont{Ryu}},
  \bibinfo{journal}{Physical Review B} \textbf{\bibinfo{volume}{94}}
  (\bibinfo{year}{2016}),
  \urlprefix\url{https://doi.org/10.1103%2Fphysrevb.94.045113}.

\bibitem[{\citenamefont{Schoutens and Wen}(2016)}]{Schoutens:2015uia}
\bibinfo{author}{\bibfnamefont{K.}~\bibnamefont{Schoutens}} \bibnamefont{and}
  \bibinfo{author}{\bibfnamefont{X.-G.} \bibnamefont{Wen}},
  \bibinfo{journal}{Phys. Rev. B} \textbf{\bibinfo{volume}{93}},
  \bibinfo{pages}{045109} (\bibinfo{year}{2016}), \eprint{1508.01111}.

\bibitem[{\citenamefont{Moore and Read}(1991)}]{Moore:1991ks}
\bibinfo{author}{\bibfnamefont{G.~W.} \bibnamefont{Moore}} \bibnamefont{and}
  \bibinfo{author}{\bibfnamefont{N.}~\bibnamefont{Read}},
  \bibinfo{journal}{Nucl. Phys. B} \textbf{\bibinfo{volume}{360}},
  \bibinfo{pages}{362} (\bibinfo{year}{1991}).

\bibitem[{\citenamefont{Ino}(1998)}]{Ino:1998by}
\bibinfo{author}{\bibfnamefont{K.}~\bibnamefont{Ino}}, \bibinfo{journal}{Nucl.
  Phys. B} \textbf{\bibinfo{volume}{532}}, \bibinfo{pages}{783}
  (\bibinfo{year}{1998}), \eprint{cond-mat/9804198}.

\bibitem[{\citenamefont{{Levin}}(2013)}]{2013PhRvX...3b1009L}
\bibinfo{author}{\bibfnamefont{M.}~\bibnamefont{{Levin}}},
  \bibinfo{journal}{Physical Review X} \textbf{\bibinfo{volume}{3}},
  \bibinfo{eid}{021009} (\bibinfo{year}{2013}), \eprint{1301.7355}.

\bibitem[{\citenamefont{Chatterjee and
  Wen}(2022{\natexlab{a}})}]{Chatterjee:2022tyg}
\bibinfo{author}{\bibfnamefont{A.}~\bibnamefont{Chatterjee}} \bibnamefont{and}
  \bibinfo{author}{\bibfnamefont{X.-G.} \bibnamefont{Wen}}
  (\bibinfo{year}{2022}{\natexlab{a}}), \eprint{2205.06244}.

\bibitem[{\citenamefont{Chatterjee and
  Wen}(2022{\natexlab{b}})}]{Chatterjee:2022kxb}
\bibinfo{author}{\bibfnamefont{A.}~\bibnamefont{Chatterjee}} \bibnamefont{and}
  \bibinfo{author}{\bibfnamefont{X.-G.} \bibnamefont{Wen}}
  (\bibinfo{year}{2022}{\natexlab{b}}), \eprint{2203.03596}.

\bibitem[{\citenamefont{Fukusumi}({\natexlab{a}})}]{Fukusumi_2022_f}
\bibinfo{author}{\bibfnamefont{Y.}~\bibnamefont{Fukusumi}}, \bibinfo{note}{in
  preparation}.

\bibitem[{\citenamefont{Nguyen et~al.}(2022)\citenamefont{Nguyen, Prabhu,
  Balram, and Gromov}}]{Nguyen:2022khy}
\bibinfo{author}{\bibfnamefont{D.~X.} \bibnamefont{Nguyen}},
  \bibinfo{author}{\bibfnamefont{K.}~\bibnamefont{Prabhu}},
  \bibinfo{author}{\bibfnamefont{A.~C.} \bibnamefont{Balram}},
  \bibnamefont{and} \bibinfo{author}{\bibfnamefont{A.}~\bibnamefont{Gromov}}
  (\bibinfo{year}{2022}), \eprint{2212.00686}.

\bibitem[{\citenamefont{Ino}(2000)}]{Ino:2000wa}
\bibinfo{author}{\bibfnamefont{K.}~\bibnamefont{Ino}} (\bibinfo{year}{2000}),
  \eprint{cond-mat/0011101}.

\bibitem[{\citenamefont{Jolicoeur et~al.}(2014)\citenamefont{Jolicoeur,
  Mizusaki, and Lecheminant}}]{Jolicoeur:2014isa}
\bibinfo{author}{\bibfnamefont{T.}~\bibnamefont{Jolicoeur}},
  \bibinfo{author}{\bibfnamefont{T.}~\bibnamefont{Mizusaki}}, \bibnamefont{and}
  \bibinfo{author}{\bibfnamefont{P.}~\bibnamefont{Lecheminant}},
  \bibinfo{journal}{Phys. Rev. B} \textbf{\bibinfo{volume}{90}},
  \bibinfo{pages}{075116} (\bibinfo{year}{2014}), \eprint{1406.5891}.

\bibitem[{\citenamefont{Estienne et~al.}(2015)\citenamefont{Estienne, Regnault,
  and Bernevig}}]{Estienne2015CorrelationLA}
\bibinfo{author}{\bibfnamefont{B.}~\bibnamefont{Estienne}},
  \bibinfo{author}{\bibfnamefont{N.}~\bibnamefont{Regnault}}, \bibnamefont{and}
  \bibinfo{author}{\bibfnamefont{B.~A.} \bibnamefont{Bernevig}},
  \bibinfo{journal}{Physical review letters} \textbf{\bibinfo{volume}{114 18}},
  \bibinfo{pages}{186801} (\bibinfo{year}{2015}).

\bibitem[{\citenamefont{Gannon}(2003)}]{Gannon:2003de}
\bibinfo{author}{\bibfnamefont{T.}~\bibnamefont{Gannon}},
  \bibinfo{journal}{Nucl. Phys. B} \textbf{\bibinfo{volume}{670}},
  \bibinfo{pages}{335} (\bibinfo{year}{2003}), \eprint{hep-th/0305070}.

\bibitem[{\citenamefont{Weerasinghe and Seidel}(2014)}]{Weerasinghe_2014}
\bibinfo{author}{\bibfnamefont{A.}~\bibnamefont{Weerasinghe}} \bibnamefont{and}
  \bibinfo{author}{\bibfnamefont{A.}~\bibnamefont{Seidel}},
  \bibinfo{journal}{Physical Review B} \textbf{\bibinfo{volume}{90}}
  (\bibinfo{year}{2014}),
  \urlprefix\url{https://doi.org/10.1103%2Fphysrevb.90.125146}.

\bibitem[{\citenamefont{Read}(2009{\natexlab{a}})}]{Read:2007cv}
\bibinfo{author}{\bibfnamefont{N.}~\bibnamefont{Read}}, \bibinfo{journal}{Phys.
  Rev. B} \textbf{\bibinfo{volume}{79}}, \bibinfo{pages}{245304}
  (\bibinfo{year}{2009}{\natexlab{a}}), \eprint{0711.0543}.

\bibitem[{\citenamefont{Read}(2009{\natexlab{b}})}]{Read:2008rn}
\bibinfo{author}{\bibfnamefont{N.}~\bibnamefont{Read}}, \bibinfo{journal}{Phys.
  Rev. B} \textbf{\bibinfo{volume}{79}}, \bibinfo{pages}{045308}
  (\bibinfo{year}{2009}{\natexlab{b}}), \eprint{0805.2507}.

\bibitem[{\citenamefont{Kedem et~al.}(1993)\citenamefont{Kedem, Klassen, McCoy,
  and Melzer}}]{Kedem:1993ze}
\bibinfo{author}{\bibfnamefont{R.}~\bibnamefont{Kedem}},
  \bibinfo{author}{\bibfnamefont{T.~R.} \bibnamefont{Klassen}},
  \bibinfo{author}{\bibfnamefont{B.~M.} \bibnamefont{McCoy}}, \bibnamefont{and}
  \bibinfo{author}{\bibfnamefont{E.}~\bibnamefont{Melzer}},
  \bibinfo{journal}{Phys. Lett. B} \textbf{\bibinfo{volume}{307}},
  \bibinfo{pages}{68} (\bibinfo{year}{1993}), \eprint{hep-th/9301046}.

\bibitem[{\citenamefont{Berkovich and McCoy}(1996)}]{Berkovich:1994es}
\bibinfo{author}{\bibfnamefont{A.}~\bibnamefont{Berkovich}} \bibnamefont{and}
  \bibinfo{author}{\bibfnamefont{B.~M.} \bibnamefont{McCoy}},
  \bibinfo{journal}{Lett. Math. Phys.} \textbf{\bibinfo{volume}{37}},
  \bibinfo{pages}{49} (\bibinfo{year}{1996}), \eprint{hep-th/9412030}.

\bibitem[{\citenamefont{Jacob and Mathieu}(2006)}]{Jacob:2005kc}
\bibinfo{author}{\bibfnamefont{P.}~\bibnamefont{Jacob}} \bibnamefont{and}
  \bibinfo{author}{\bibfnamefont{P.}~\bibnamefont{Mathieu}},
  \bibinfo{journal}{Nucl. Phys. B} \textbf{\bibinfo{volume}{733}},
  \bibinfo{pages}{205} (\bibinfo{year}{2006}), \eprint{hep-th/0506074}.

\bibitem[{\citenamefont{Hsieh et~al.}(2020)\citenamefont{Hsieh, Nakayama, and
  Tachikawa}}]{Hsieh:2020uwb}
\bibinfo{author}{\bibfnamefont{C.-T.} \bibnamefont{Hsieh}},
  \bibinfo{author}{\bibfnamefont{Y.}~\bibnamefont{Nakayama}}, \bibnamefont{and}
  \bibinfo{author}{\bibfnamefont{Y.}~\bibnamefont{Tachikawa}}
  (\bibinfo{year}{2020}), \eprint{2002.12283}.

\bibitem[{\citenamefont{Kulp}(2020)}]{Kulp:2020iet}
\bibinfo{author}{\bibfnamefont{J.}~\bibnamefont{Kulp}} (\bibinfo{year}{2020}),
  \eprint{2003.04278}.

\bibitem[{\citenamefont{Lou et~al.}(2021)\citenamefont{Lou, Shen, Chen, and
  Hung}}]{Lou:2020gfq}
\bibinfo{author}{\bibfnamefont{J.}~\bibnamefont{Lou}},
  \bibinfo{author}{\bibfnamefont{C.}~\bibnamefont{Shen}},
  \bibinfo{author}{\bibfnamefont{C.}~\bibnamefont{Chen}}, \bibnamefont{and}
  \bibinfo{author}{\bibfnamefont{L.-Y.} \bibnamefont{Hung}},
  \bibinfo{journal}{JHEP} \textbf{\bibinfo{volume}{02}}, \bibinfo{pages}{171}
  (\bibinfo{year}{2021}), \eprint{2007.10562}.

\bibitem[{\citenamefont{Inamura}(2022)}]{Inamura:2022lun}
\bibinfo{author}{\bibfnamefont{K.}~\bibnamefont{Inamura}}
  (\bibinfo{year}{2022}), \eprint{2206.13159}.

\bibitem[{\citenamefont{Creutzig and Ridout}(2013)}]{Creutzig:2013hma}
\bibinfo{author}{\bibfnamefont{T.}~\bibnamefont{Creutzig}} \bibnamefont{and}
  \bibinfo{author}{\bibfnamefont{D.}~\bibnamefont{Ridout}},
  \bibinfo{journal}{J. Phys. A} \textbf{\bibinfo{volume}{46}},
  \bibinfo{pages}{4006} (\bibinfo{year}{2013}), \eprint{1303.0847}.

\bibitem[{\citenamefont{Guruswamy and Ludwig}(1998)}]{Guruswamy:1996rk}
\bibinfo{author}{\bibfnamefont{S.}~\bibnamefont{Guruswamy}} \bibnamefont{and}
  \bibinfo{author}{\bibfnamefont{A.~W.~W.} \bibnamefont{Ludwig}},
  \bibinfo{journal}{Nucl. Phys. B} \textbf{\bibinfo{volume}{519}},
  \bibinfo{pages}{661} (\bibinfo{year}{1998}), \eprint{hep-th/9612172}.

\bibitem[{\citenamefont{Bernevig and Haldane}(2008)}]{Bernevig_2008}
\bibinfo{author}{\bibfnamefont{B.~A.} \bibnamefont{Bernevig}} \bibnamefont{and}
  \bibinfo{author}{\bibfnamefont{F.~D.~M.} \bibnamefont{Haldane}},
  \bibinfo{journal}{Physical Review Letters} \textbf{\bibinfo{volume}{100}}
  (\bibinfo{year}{2008}),
  \urlprefix\url{https://doi.org/10.1103%2Fphysrevlett.100.246802}.

\bibitem[{\citenamefont{Bernevig et~al.}(2009)\citenamefont{Bernevig, Gurarie,
  and Simon}}]{Bernevig_2009}
\bibinfo{author}{\bibfnamefont{B.~A.} \bibnamefont{Bernevig}},
  \bibinfo{author}{\bibfnamefont{V.}~\bibnamefont{Gurarie}}, \bibnamefont{and}
  \bibinfo{author}{\bibfnamefont{S.~H.} \bibnamefont{Simon}},
  \bibinfo{journal}{Journal of Physics A: Mathematical and Theoretical}
  \textbf{\bibinfo{volume}{42}}, \bibinfo{pages}{245206}
  (\bibinfo{year}{2009}),
  \urlprefix\url{https://doi.org/10.1088%2F1751-8113%2F42%2F24%2F245206}.

\bibitem[{\citenamefont{Milovanovic and Read}(1996)}]{Milovanovic:1996nj}
\bibinfo{author}{\bibfnamefont{M.}~\bibnamefont{Milovanovic}} \bibnamefont{and}
  \bibinfo{author}{\bibfnamefont{N.}~\bibnamefont{Read}},
  \bibinfo{journal}{Phys. Rev. B} \textbf{\bibinfo{volume}{53}},
  \bibinfo{pages}{13559} (\bibinfo{year}{1996}), \eprint{cond-mat/9602113}.

\bibitem[{\citenamefont{Alcaraz et~al.}(1988)\citenamefont{Alcaraz, Barber, and
  Batchelor}}]{Alcaraz:1987zr}
\bibinfo{author}{\bibfnamefont{F.~C.} \bibnamefont{Alcaraz}},
  \bibinfo{author}{\bibfnamefont{M.~N.} \bibnamefont{Barber}},
  \bibnamefont{and} \bibinfo{author}{\bibfnamefont{M.~T.}
  \bibnamefont{Batchelor}}, \bibinfo{journal}{Annals Phys.}
  \textbf{\bibinfo{volume}{182}}, \bibinfo{pages}{280} (\bibinfo{year}{1988}).

\bibitem[{\citenamefont{Kitazawa}(1997)}]{Kitazawa_1997}
\bibinfo{author}{\bibfnamefont{A.}~\bibnamefont{Kitazawa}},
  \bibinfo{journal}{Journal of Physics A: Mathematical and General}
  \textbf{\bibinfo{volume}{30}}, \bibinfo{pages}{L285} (\bibinfo{year}{1997}),
  \urlprefix\url{https://doi.org/10.1088%2F0305-4470%2F30%2F9%2F005}.

\bibitem[{\citenamefont{Nomura and Kitazawa}(1998)}]{Nomura:1997pp}
\bibinfo{author}{\bibfnamefont{K.}~\bibnamefont{Nomura}} \bibnamefont{and}
  \bibinfo{author}{\bibfnamefont{A.}~\bibnamefont{Kitazawa}},
  \bibinfo{journal}{J. Phys. A} \textbf{\bibinfo{volume}{31}},
  \bibinfo{pages}{7341} (\bibinfo{year}{1998}), \eprint{cond-mat/9711294}.

\bibitem[{\citenamefont{Thorngren}(2020)}]{Thorngren:2018bhj}
\bibinfo{author}{\bibfnamefont{R.}~\bibnamefont{Thorngren}},
  \bibinfo{journal}{Commun. Math. Phys.} \textbf{\bibinfo{volume}{378}},
  \bibinfo{pages}{1775} (\bibinfo{year}{2020}), \eprint{1810.04414}.

\bibitem[{\citenamefont{Yao and Fukusumi}(2019)}]{Yao:2019bub}
\bibinfo{author}{\bibfnamefont{Y.}~\bibnamefont{Yao}} \bibnamefont{and}
  \bibinfo{author}{\bibfnamefont{Y.}~\bibnamefont{Fukusumi}},
  \bibinfo{journal}{Phys. Rev. B} \textbf{\bibinfo{volume}{100}},
  \bibinfo{pages}{075105} (\bibinfo{year}{2019}), \eprint{1902.06584}.

\bibitem[{\citenamefont{Picco et~al.}(2016)\citenamefont{Picco, Ribault, and
  Santachiara}}]{Picco:2016ilr}
\bibinfo{author}{\bibfnamefont{M.}~\bibnamefont{Picco}},
  \bibinfo{author}{\bibfnamefont{S.}~\bibnamefont{Ribault}}, \bibnamefont{and}
  \bibinfo{author}{\bibfnamefont{R.}~\bibnamefont{Santachiara}},
  \bibinfo{journal}{SciPost Phys.} \textbf{\bibinfo{volume}{1}},
  \bibinfo{pages}{009} (\bibinfo{year}{2016}), \eprint{1607.07224}.

\bibitem[{\citenamefont{Gori and Viti}(2018)}]{Gori:2018gqx}
\bibinfo{author}{\bibfnamefont{G.}~\bibnamefont{Gori}} \bibnamefont{and}
  \bibinfo{author}{\bibfnamefont{J.}~\bibnamefont{Viti}},
  \bibinfo{journal}{JHEP} \textbf{\bibinfo{volume}{12}}, \bibinfo{pages}{131}
  (\bibinfo{year}{2018}), \eprint{1806.02330}.

\bibitem[{\citenamefont{Picco et~al.}(2019)\citenamefont{Picco, Ribault, and
  Santachiara}}]{Picco:2019dkm}
\bibinfo{author}{\bibfnamefont{M.}~\bibnamefont{Picco}},
  \bibinfo{author}{\bibfnamefont{S.}~\bibnamefont{Ribault}}, \bibnamefont{and}
  \bibinfo{author}{\bibfnamefont{R.}~\bibnamefont{Santachiara}},
  \bibinfo{journal}{SciPost Phys.} \textbf{\bibinfo{volume}{7}},
  \bibinfo{pages}{044} (\bibinfo{year}{2019}), \eprint{1906.02566}.

\bibitem[{\citenamefont{Lykke~Jacobsen and
  Saleur}(2019)}]{LykkeJacobsen:2018cbt}
\bibinfo{author}{\bibfnamefont{J.}~\bibnamefont{Lykke~Jacobsen}}
  \bibnamefont{and} \bibinfo{author}{\bibfnamefont{H.}~\bibnamefont{Saleur}},
  \bibinfo{journal}{JHEP} \textbf{\bibinfo{volume}{01}}, \bibinfo{pages}{084}
  (\bibinfo{year}{2019}), \eprint{1809.02191}.

\bibitem[{\citenamefont{Harvey et~al.}(2020)\citenamefont{Harvey, Hu, and
  Wu}}]{Harvey:2019qzs}
\bibinfo{author}{\bibfnamefont{J.~A.} \bibnamefont{Harvey}},
  \bibinfo{author}{\bibfnamefont{Y.}~\bibnamefont{Hu}}, \bibnamefont{and}
  \bibinfo{author}{\bibfnamefont{Y.}~\bibnamefont{Wu}}, \bibinfo{journal}{J.
  Phys. A} \textbf{\bibinfo{volume}{53}}, \bibinfo{pages}{334003}
  (\bibinfo{year}{2020}), \eprint{1912.11955}.

\bibitem[{\citenamefont{Runkel and Watts}(2020)}]{Runkel:2020zgg}
\bibinfo{author}{\bibfnamefont{I.}~\bibnamefont{Runkel}} \bibnamefont{and}
  \bibinfo{author}{\bibfnamefont{G.~M.~T.} \bibnamefont{Watts}},
  \bibinfo{journal}{JHEP} \textbf{\bibinfo{volume}{06}}, \bibinfo{pages}{025}
  (\bibinfo{year}{2020}), \eprint{2001.05055}.

\bibitem[{\citenamefont{Makabe and Watts}(2017)}]{Makabe:2017ygy}
\bibinfo{author}{\bibfnamefont{I.}~\bibnamefont{Makabe}} \bibnamefont{and}
  \bibinfo{author}{\bibfnamefont{G.~M.~T.} \bibnamefont{Watts}},
  \bibinfo{journal}{JHEP} \textbf{\bibinfo{volume}{09}}, \bibinfo{pages}{013}
  (\bibinfo{year}{2017}), \eprint{1703.09148}.

\bibitem[{\citenamefont{Hao et~al.}(2022)\citenamefont{Hao, Li, and
  Qi}}]{Hao:2022kxo}
\bibinfo{author}{\bibfnamefont{J.-X.} \bibnamefont{Hao}},
  \bibinfo{author}{\bibfnamefont{W.}~\bibnamefont{Li}}, \bibnamefont{and}
  \bibinfo{author}{\bibfnamefont{Y.}~\bibnamefont{Qi}} (\bibinfo{year}{2022}),
  \eprint{2211.05618}.

\bibitem[{\citenamefont{Gaiotto and Kapustin}(2016)}]{Gaiotto:2015zta}
\bibinfo{author}{\bibfnamefont{D.}~\bibnamefont{Gaiotto}} \bibnamefont{and}
  \bibinfo{author}{\bibfnamefont{A.}~\bibnamefont{Kapustin}},
  \bibinfo{journal}{Int. J. Mod. Phys. A} \textbf{\bibinfo{volume}{31}},
  \bibinfo{pages}{1645044} (\bibinfo{year}{2016}), \eprint{1505.05856}.

\bibitem[{\citenamefont{Cappelli and Viola}(2011)}]{Cappelli:2010jv}
\bibinfo{author}{\bibfnamefont{A.}~\bibnamefont{Cappelli}} \bibnamefont{and}
  \bibinfo{author}{\bibfnamefont{G.}~\bibnamefont{Viola}}, \bibinfo{journal}{J.
  Phys. A} \textbf{\bibinfo{volume}{44}}, \bibinfo{pages}{075401}
  (\bibinfo{year}{2011}), \eprint{1007.1732}.

\bibitem[{\citenamefont{Papi{\'{c} }}(2014)}]{Papi__2014}
\bibinfo{author}{\bibfnamefont{Z.}~\bibnamefont{Papi{\'{c} }}},
  \bibinfo{journal}{Physical Review B} \textbf{\bibinfo{volume}{90}}
  (\bibinfo{year}{2014}),
  \urlprefix\url{https://doi.org/10.1103%2Fphysrevb.90.075304}.

\bibitem[{\citenamefont{Tachikawa}(2020)}]{Tachikawa:2017gyf}
\bibinfo{author}{\bibfnamefont{Y.}~\bibnamefont{Tachikawa}},
  \bibinfo{journal}{SciPost Phys.} \textbf{\bibinfo{volume}{8}},
  \bibinfo{pages}{015} (\bibinfo{year}{2020}), \eprint{1712.09542}.

\bibitem[{\citenamefont{Bhardwaj and Tachikawa}(2018)}]{Bhardwaj:2017xup}
\bibinfo{author}{\bibfnamefont{L.}~\bibnamefont{Bhardwaj}} \bibnamefont{and}
  \bibinfo{author}{\bibfnamefont{Y.}~\bibnamefont{Tachikawa}},
  \bibinfo{journal}{JHEP} \textbf{\bibinfo{volume}{03}}, \bibinfo{pages}{189}
  (\bibinfo{year}{2018}), \eprint{1704.02330}.

\bibitem[{\citenamefont{Chang et~al.}(2022)\citenamefont{Chang, Chen, and
  Xu}}]{Chang:2022hud}
\bibinfo{author}{\bibfnamefont{C.-M.} \bibnamefont{Chang}},
  \bibinfo{author}{\bibfnamefont{J.}~\bibnamefont{Chen}}, \bibnamefont{and}
  \bibinfo{author}{\bibfnamefont{F.}~\bibnamefont{Xu}} (\bibinfo{year}{2022}),
  \eprint{2208.02757}.

\bibitem[{\citenamefont{Petkova and Zuber}(2001)}]{Petkova:2000ip}
\bibinfo{author}{\bibfnamefont{V.~B.} \bibnamefont{Petkova}} \bibnamefont{and}
  \bibinfo{author}{\bibfnamefont{J.~B.} \bibnamefont{Zuber}},
  \bibinfo{journal}{Phys. Lett. B} \textbf{\bibinfo{volume}{504}},
  \bibinfo{pages}{157} (\bibinfo{year}{2001}), \eprint{hep-th/0011021}.

\bibitem[{\citenamefont{Fukusumi}({\natexlab{b}})}]{Fukusumi_2022_c}
\bibinfo{author}{\bibfnamefont{Y.}~\bibnamefont{Fukusumi}}, \bibinfo{note}{in
  preparation}.

\bibitem[{\citenamefont{Lu}(1989)}]{LU198946}
\bibinfo{author}{\bibfnamefont{S.}~\bibnamefont{Lu}}, \bibinfo{journal}{Physics
  Letters B} \textbf{\bibinfo{volume}{218}}, \bibinfo{pages}{46}
  (\bibinfo{year}{1989}), ISSN \bibinfo{issn}{0370-2693},
  \urlprefix\url{https://www.sciencedirect.com/science/article/pii/0370269389904723}.

\bibitem[{\citenamefont{Jacobsen et~al.}(2022)\citenamefont{Jacobsen, Ribault,
  and Saleur}}]{Jacobsen:2022nxs}
\bibinfo{author}{\bibfnamefont{J.~L.} \bibnamefont{Jacobsen}},
  \bibinfo{author}{\bibfnamefont{S.}~\bibnamefont{Ribault}}, \bibnamefont{and}
  \bibinfo{author}{\bibfnamefont{H.}~\bibnamefont{Saleur}}
  (\bibinfo{year}{2022}), \eprint{2208.14298}.

\bibitem[{\citenamefont{Nakamura et~al.}(2012)\citenamefont{Nakamura, Wang, and
  Bergholtz}}]{Nakamura_2012}
\bibinfo{author}{\bibfnamefont{M.}~\bibnamefont{Nakamura}},
  \bibinfo{author}{\bibfnamefont{Z.-Y.} \bibnamefont{Wang}}, \bibnamefont{and}
  \bibinfo{author}{\bibfnamefont{E.~J.} \bibnamefont{Bergholtz}},
  \bibinfo{journal}{Physical Review Letters} \textbf{\bibinfo{volume}{109}}
  (\bibinfo{year}{2012}),
  \urlprefix\url{https://doi.org/10.1103%2Fphysrevlett.109.016401}.

\bibitem[{\citenamefont{Bernevig and
  Regnault}(2012)}]{https://doi.org/10.48550/arxiv.1204.5682}
\bibinfo{author}{\bibfnamefont{B.~A.} \bibnamefont{Bernevig}} \bibnamefont{and}
  \bibinfo{author}{\bibfnamefont{N.}~\bibnamefont{Regnault}},
  \emph{\bibinfo{title}{Thin-torus limit of fractional topological insulators}}
  (\bibinfo{year}{2012}), \urlprefix\url{https://arxiv.org/abs/1204.5682}.

\bibitem[{\citenamefont{Faddeev}(1984)}]{Faddeev:1984jp}
\bibinfo{author}{\bibfnamefont{L.~D.} \bibnamefont{Faddeev}},
  \bibinfo{journal}{Phys. Lett. B} \textbf{\bibinfo{volume}{145}},
  \bibinfo{pages}{81} (\bibinfo{year}{1984}).

\bibitem[{\citenamefont{Hosono}(1989)}]{Hosono:1987na}
\bibinfo{author}{\bibfnamefont{S.}~\bibnamefont{Hosono}},
  \bibinfo{journal}{Phys. Rev. D} \textbf{\bibinfo{volume}{39}},
  \bibinfo{pages}{1743} (\bibinfo{year}{1989}).

\bibitem[{\citenamefont{Fujiwara et~al.}(1988)\citenamefont{Fujiwara, Hosono,
  and Kitakado}}]{Fujiwara:1988zc}
\bibinfo{author}{\bibfnamefont{T.}~\bibnamefont{Fujiwara}},
  \bibinfo{author}{\bibfnamefont{S.}~\bibnamefont{Hosono}}, \bibnamefont{and}
  \bibinfo{author}{\bibfnamefont{S.}~\bibnamefont{Kitakado}},
  \bibinfo{journal}{Mod. Phys. Lett. A} \textbf{\bibinfo{volume}{3}},
  \bibinfo{pages}{1585} (\bibinfo{year}{1988}).

\end{thebibliography}

\end{document}